\documentclass[10pt,journal,compsoc]{IEEEtran}

\ifCLASSOPTIONcompsoc
  \usepackage[nocompress]{cite}
\else
  \usepackage{cite}
\fi

\usepackage{cite}
\usepackage{amsmath,amssymb,amsfonts,amsthm}
\usepackage[ruled]{algorithm2e}
\usepackage{graphicx}
\usepackage{subfig}
\usepackage{textcomp}
\usepackage{color}
\usepackage{booktabs}
\usepackage{multirow}
\usepackage{url}
\newtheorem{theorem}{\textbf{Lemma}}

\newcommand{\js}{\textcolor{black}}
\newcommand{\jsnew}{\textcolor{black}}
\newcommand{\wjs}{\textcolor{black}}
\newcommand{\wjss}{\textcolor{black}}
\newcommand{\wjst}{\textcolor{black}}
\usepackage{ragged2e}
\usepackage{balance}
\usepackage{tikz}
\newcommand*{\circled}[1]{\lower.7ex\hbox{\tikz\draw (0pt, 0pt)%
    circle (.5em) node {\makebox[.6em][c]{\small #1}};}}

\usepackage{xcolor}
\usepackage{xpatch}

\makeatletter
\def\changeBibColor#1{%
  \in@{#1}{bunz2020zether, hoekstra2013using, pass2017formal, Kaggle, Ben14, benhamouda2019supporting, parno2013pinocchio, das2019fastkitten, chen2019opera, liu2020confidential, orenbach2017eleos, orenbach2019cosmix, gretton2012kernel, anderson1994two, yang2018lightweight}
  \ifin@\color{black}\else\normalcolor\fi
}

\xpatchcmd\@bibitem
  {\item}
  {\changeBibColor{#1}\item}
  {}{\fail}

\xpatchcmd\@lbibitem
  {\item}
  {\changeBibColor{#2}\item}
  {}{\fail}
\makeatother

\begin{document}

\title{Golden Grain: Building a Secure and Decentralized Model Marketplace for MLaaS}

\author{Jiasi Weng${}^*$,~Jian Weng,~\IEEEmembership{Member,~IEEE},~Chengjun Cai, Hongwei Huang, Cong Wang,~\IEEEmembership{Fellow,~IEEE} 
\IEEEcompsocitemizethanks{

\IEEEcompsocthanksitem
Jiasi Weng, Jian Weng and Hongwei Huang are with the College of Information Science and Technology / College of Cyber security, National Joint Engineering Research Center of Network Security Detection and Protection Technology, and Guangdong Key Laboratory of Data Security and Privacy Preserving, Jinan University, Guangzhou, 510632, China. Jiasi Weng is also with the Department of Computer Science, City University of Hong Kong, Hong Kong. E-mail: cryptjweng@gmail.com; hwhuang001e@outlook.com; wengjiasi@gmail.com.
\protect\\
Chengjun Cai and Cong Wang are with the Department of Computer Science, City University of Hong Kong,  Hong Kong, and also with the City University of Hong Kong Shenzhen Research Institute, Shenzhen 518057, China.
E-mail: chencai-c@my.cityu.edu.hk; congwang@cityu.edu.hk.
\protect\\
Jian Weng is the corresponding author: cryptjweng@gmail.com.
\protect\\
${}^*$ Work was done when the author was a research assistant in the City University of Hong Kong.
}
}

\IEEEtitleabstractindextext{%
\begin{abstract}
\justifying
ML-as-a-service (MLaaS) becomes increasingly popular and revolutionizes the lives of people.
A natural requirement for MLaaS is, however, to provide highly accurate prediction services.
To achieve this, current MLaaS systems integrate and combine multiple well-trained models in their services.
Yet, in reality, there is no easy way for MLaaS providers, especially for startups, to collect sufficiently well-trained models from individual developers, due to the lack of incentives.
In this paper, we aim to fill this gap by building up a model marketplace, called as Golden Grain, to facilitate model sharing, which enforces the fair model-money swapping process between individual developers and MLaaS providers.
Specifically, we deploy the swapping process on the blockchain, and further introduce a blockchain-empowered model benchmarking process for transparently determining the model prices according to their authentic performances, so as to motivate the faithful contributions of well-trained models.
Especially, to ease the blockchain overhead for model benchmarking, our marketplace carefully offloads the heavy computation and designs a secure off-chain on-chain interaction protocol based on a trusted execution environment (TEE), for ensuring both the integrity and authenticity of benchmarking.
We implement a prototype of our Golden Grain on the Ethereum blockchain, and conduct extensive experiments using standard benchmark datasets to demonstrate the practically affordable performance of our design.
\end{abstract}

}

\maketitle

\IEEEraisesectionheading{
\section{INTRODUCTION}\label{intro}}
Machine learning (ML) systems and algorithms have fundamentally revolutionized the lives of people with various ML-based applications, such as face recognition~\cite{ahonen2006face}, automated vehicles~\cite{ohn2016looking} and disease diagnosis~\cite{lecun2015deep, danaee2017deep}.
\js{Among others, one important improvement for ML is ML-as-a-service (MLaaS)~\cite{ribeiro2015mlaas}, leaded by giants, \emph{e.g.}, Google~\cite{Google}, Amazon~\cite{AMAZON} and Microsoft~\cite{MICROSOFT}, and supported by growing startup companies~\cite{Statista}.}
MLaaS makes ML practical for everyone by easing users' need for training models before they can enjoy ML services.
While MLaaS greatly facilitates users, it brings very critical requirements for MLaaS providers.
For example, it is natural to ensure that the predictions given from the services should be as accurate as possible~\cite{crankshaw2017clipper, lee2018pretzel, olston2017tensorflow}.
Toward this goal, recently several prediction serving systems like Clipper~\cite{crankshaw2017clipper} and PRETZEL~\cite{lee2018pretzel} are proposed, which integrate and merge predictions from multiple models, with the aim of obtaining the better prediction accuracy and robustness than TensorFlow Serving~\cite{olston2017tensorflow}.~\looseness=-1

Despite being very promising, they usually assume that the trained models used for accurate prediction serving are collected as a prior and that they are well-trained (\textit{i.e.}, they have good model accuracy).
\js{However, in reality an MLaaS provider, especially a startup company, faces a lot of challenges to securely collect well-trained models.}
Firstly, large companies, such as Google and Amazon, might be reluctant to release their models which are trained on sensitive user data, due to data protection regulations~\cite{goddard2017eu}.
Individual developers (or model owners) may also be unwilling to publicly share their trained models, since pre-trained models memorize the data they train on, and thus the models could reveal private information about the training data via recent membership inference attacks~\cite{shokri2017membership} and model extraction attacks~\cite{tramer2016stealing, carlini2018secret}.
Secondly, many MLaaS startups have no access to an experienced and qualified data science team for training good models~\cite{LeanAI}.
Although they can turn to individual developers for help, the quality of the trained models can hardly be enforced in case of moving them from experimentation to production, especially when we further consider the negative effects of shifted data and adversarial examples~\cite{hendrycks2019benchmarking, hendrycks2019augmix}.

On the other hand, we are aware that existing data marketplaces, such as Sterling~\cite{hynes2018demonstration} and OpenMined~\cite{Openmined}, have paved the way for training models by enabling secure data sharing, which eliminates the obstacle of obtaining available training data.
\wjss{Unfortunately, there still lacks a connection between collecting trained models and realizing robust predictions for MLaaS, due to the lack of secure mechanisms for monetizing the trained models.}
Therefore, the above issues inspire this work: \emph{We build a model marketplace to bridge model developers and MLaaS providers, which leverages incentives to facilitate model sharing and motivate the faithful contributions of well-trained models, and in the meantime, fully respects the model privacy.}~\looseness=-1

Building such a model marketplace, however, is a non-trivial task.
To start with, it is easily understood that fairness between the model seller \wjss{(\emph{e.g.}, model owner)} and the model buyer (\emph{e.g.}, MLaaS provider) should be enforced~\cite{li2018crowdbc, cai2018enabling}.
That is, a seller should reveal nothing about its model unless someone pays to purchase the model, and meanwhile, the buyer should indeed obtain the model he is interested in once the money is spent.
Usually, fulfilling such a fair swapping process would require a trusted third party (TTP) to enforce fairness of the process and resolve disputes~\cite{pagnia1999impossibility}.
But using such a TTP would bring several disadvantages, \emph{e.g.}, lack of process transparency~\cite{lu2018zebralancer, li2019toward}.
To build a fair marketplace without a TTP, blockchain is a widely adopted choice for its transparency and execution correctness~\cite{kosba2016hawk, tramer2017sealed, cai2019building}.\looseness=-1

Besides, we also need a secure mechanism to incentivize the model sellers to provide well-trained models for the marketplace, but how to fulfill our goal satisfactorily?
Particularly, we have to address the following two challenging problems.

\noindent\textbf{Challenge 1: correct model benchmarking while protecting model privacy.}
To incentivize the sellers to provide as good as possible models for the marketplace,
\js{
a natural idea is to determine a model's price in proportion to the model's authentic performance, where the model performance can be measured via benchmarking on unrevealed datasets, by following a consensus in ML practice~\cite{hendrycks2019benchmarking, hendrycks2019augmix}.
By this way, sellers might be motivated to provide better models for earning more money~\cite{kranton2003competition}.
}
At a first glance, \js{if we are concerned about a malicious seller,} it might appear that we can readily put his/her model on-chain, and run the benchmarking process through crafting smart contract to correctly record the result performance and determine its price in later monetization.
\emph{But such an approach is inefficient, since the model might be large, so readily storing it on-chain would incur unaffordable costs.
Besides, putting the model on-chain would also compromise the model privacy, since the blockchain is publicly accessible.}

\noindent\textbf{Challenge 2: authenticated benchmark dataset relaying.}
For guaranteeing the correctness of the benchmarking process, we also need an unbiased benchmark dataset for correctly calculating each model's performance.
This is, however, challenging because although there are reputable benchmark datasets available for benchmarking, \emph{e.g.}, \wjs{provided by authorized organizations}~\cite{hendrycks2019benchmarking, hendrycks2019augmix, ford2019adversarial}, they are stored outside the blockchain.
\emph{Requested external benchmark datasets might be tampered before being written on the blockchain, which could further result in bias for model benchmarking.
Thus, an efficient and secure mechanism for authentically relaying those requested benchmark datasets is desired.}

\noindent\textbf{Our Result.}
To address the two challenges above, our design follows a widely adopted off-chain processing approach and further resorts to a trusted execution environment (TEE) to complete \emph{\textbf{a correct model benchmarking process}} (Section~\ref{sec:roadmap_bench}).
In particular, the correctness of the model benchmarking is two-fold.
First, the benchmarking process is executed with integrity guarantees.
A seller is required to first commit to his model into the locally equipped TEE before receiving the benchmark samples, which prevents the seller from \emph{forging} a fake model during the execution or providing an unauthentic benchmarked result.
\js{
Second, the model is benchmarked by the authentically relayed benchmark samples.
Following a trend practice~\cite{zhang2016town}, in our case, the ready-made TEE-based infrastructure residing on the seller side can help relaying the required samples and recording their digests on the blockchain, so that an end-to-end authenticated channel for transmitting benchmark samples can be established.
}
Notably, our solution using TEE can be more performant than those using verifiable cryptographic tools~\cite{parno2013pinocchio}.\looseness=-1

\js{With the evaluated authentic performance of a model, our marketplace transparently decides the model price on smart contract. 
Additionally, our pricing mechanism simultaneously maximizes the revenue of sellers and the utility of buyers, in order to maintain the sustainability of our model marketplace in a long-term perspective.}
Finally, the model performance and price are correctly recorded on the blockchain.
After the model benchmarking stage, a buyer can choose a model according to the recorded information \wjss{(\emph{i.e.}, model information, performance and price)} on the blockchain and trade with the model seller via \emph{\textbf{a fair model-money swapping process}} (Section~\ref{sec:roadmap_mone}) that is built by integrating the smart contract and the TEE.
Here, the buyer will get the correct model which is previously committed to the blockchain during the benchmarking stage, if only he faithfully pays the seller on the blockchain.

\wjss{By using the widely adopted Universal Composability (UC) security model, we design two on-chain off-chain interaction protocols \emph{\textbf{to achieve the correct model benchmarking process and the fair model-money swapping process}}, for the model benchmarking stage and the model monetization stage, respectively.
To be specific, we adopt the UC-based security definitions for smart contract which are formalized by Hawk~\cite{kosba2016hawk} and the definitions for transparent TEE which are formalized in the SGP protocol~\cite{tramer2017sealed}.
Besides, we prove that our defined protocols ensure two desirable security properties including the correctness of model benchmarking and the fairness of model exchanging under the UC security framework.}

In conclusion, the main contributions of our work are as follows:
\vspace{-0.4em}
\begin{itemize}
\item We build a secure and decentralized model marketplace to facilitate model sharing via achieving the fair model-money swapping process, where the seller obtains the money \textit{iff} the buyer can correctly and timely get the authentic model of his interest.

\item \js{We design a blockchain-empowered model benchmarking process, where
model performances are correctly evaluated on authenticated \jsnew{test} datasets, and model prices are determined based on the authentic model performances, through which our marketplace motivates the faithful contributions of well-trained models. }

\item We implement a prototype of our marketplace design on the Ethereum blockchain, and the extensive experiments with three standard datasets demonstrate the practically affordable performance of our design.
\end{itemize}

\section{RELATED WORK}
\noindent\textbf{Decentralized Data Marketplace for Machine Learning.}
Decentralized data marketplaces which help connect participants to contribute reliable data for machine learning pipelines have attracted wide attention from both academia and industry, such as Sterling~\cite{hynes2018demonstration}, DATABRIGHT~\cite{dao2018databright}, DeepChain~\cite{weng2019deepchain} and OpenMined~\cite{Openmined}.
Among those mentioned designs, participants' contributed data will be protected via encryption (\emph{e.g.}, homomorphic encryption) or trusted hardware, and they will later obtain enforced rewards with respect to the quality of the contributed data via the blockchain, aiming to motivate trustworthy behaviors~\cite{li2018crowdbc, cai2018leveraging}.
Although decentralized data marketplaces shed lights on how to overcome the dilemma of data scarcity and quality in the machine learning community, we consider that they focus on collecting high-quality data for training models and lack connections to realize a robust prediction service in current MLaaS systems.
Different from them, we focus on filling this important gap by building a decentralized model marketplace \wjs{towards} robust prediction service in MLaaS systems, motivating good-quality models and enabling the fair monetization through the blockchain.~\looseness=-1

\noindent\textbf{TEE-Based Model Evaluation.}
An ML pipeline generally involves two phases, \emph{i.e.}, model training and model evaluation, and a prediction serving system mainly focuses on the latter phase.
For model privacy and integrity, model evaluation can be run by employing TEE, \emph{e.g.}, Intel's Software Guard Extensions (SGX)~\cite{mckeen2016intel}.
Recently, there exists a series of work on running the phase of model evaluation inside the TEE.
Slalom~\cite{tramer2018slalom}, in order to improve efficiency, does not run the entire phase of model evaluation inside the TEE which is only supported by CPUs.
Instead, it executes a few of non-linear computation inside the TEE and delegates all linear computation to a co-located GPU processor combining with secure primitives.
By adopting this method, it achieves more efficient evaluation, compared to the case of solely running in the TEE.
In our case, we consider that the setting of TEE coexisted with GPUs may bring constraints to sellers entering the marketplace.
PRIVADO~\cite{tople2018privado} particularly considers access pattern based attacks on the TEE and leverage the oblivious using tools to transform any deep learning model to have data-independent access patterns, eliminating the data leakage inside the TEE.
In this work, the benchmark samples are non-private during the benchmarking stage, and thus access pattern attacks on the TEE need not be considered.
Different from the aforementioned two works only focusing on the model evaluation, Chiron~\cite{hunt2018chiron} runs both phases of model training and model evaluation.
It builds a system using the TEE and Ryoan sandbox to protect data privacy and model privacy, but it has a larger Trusted Computing Base (TCB) than the mentioned two works.
In our work, we pay the main attention to securely conduct the phase of model evaluation by resorting to the TEE.

\section{PRELIMINARIES}
\wjss{Here we are ready to introduce the used cryptographic primitives, smart contract and TEE which are employed in our design.
Note that we make black-box use of the cryptographic primitives and assume that they are secure.}
\subsection{Cryptographic Primitives}
\noindent\textbf{Commitment Schemes.}
\wjss{A commitment scheme is generally denoted as $\textsf{COM}=(\textsf{Setup}, \textsf{Commit}, \textsf{Opening})$.
Specifically, \textsf{Setup} is used to generate public parameters $pub$.
\textsf{Commit} takes the public parameters $pub$, a message $msg$ and random coins $r$ as inputs and outputs a commitment to the message, formally denoted as $com \leftarrow \textsf{Commit}(pub, msg, r)$.
\textsf{Opening} is used to re-compute the commitment by using the same message $msg'$ and the unrevealed random coins $r'$; if $com=\textsf{Commit}(pub, msg', r')$, $(com, msg, r)$ is valid.}

\noindent\textbf{Zero-Knowledge Proofs.}
\wjss{A zero-knowledge proof protocol for a statement allows a potentially malicious prover to convince a honest verifier that the statement is valid without revealing any other information.
This paper realizes zero-knowledge proofs of knowledge by resorting to trusted hardware based proofs~\cite{tramer2017sealed}.}
\subsection{Smart Contracts}
\wjss{Smart contracts are Turing-complete programs which are automatically executed on the blockchain.
One of the most popular systems implementing smart contract is Ethereum.
Ethereum users can write and deploy smart contracts for performing designated business logic, defined as a sequence of entry points.
A deployed smart contract with the entry points will be invoked by messages (\emph{i.e.}, transactions) sent from other contracts or non-contract users.
Once the smart contract is deployed on the Ethereum, no one can modify it.}\looseness=-1
%

\wjss{Generally, blockchain endows smart contracts with the integrity property, but smart contracts also inherit the transparency of blockchain.
On one hand, smart contracts running on the blockchain ensure the integrity of the program execution, and thus enforce trust among distrustful parties.
On the other hand, data stored on the smart contracts are public without confidentiality guarantees.
To overcome the undesirable property, a few smart contract systems~\cite{kosba2016hawk, cheng2019ekiden, bunz2020zether} propose privacy-preserving smart contracts.}

\wjss{On Ethereum, the execution of each smart contract (and each transaction) associates with \textsf{Gas} (\emph{i.e.}, money) costs.
\textsf{Gas} consumption of a smart contract depends on the computation steps and storage space this smart contract requires.
Therefore, the successful execution of a smart contract needs amount of available \textsf{Gas}, since \textsf{Gas} limitation will restrict its execution.
Besides, the \textsf{Gas} cost can bound the computation steps of a smart contract, thereby preventing DoS attacks.
For example, a smart contract creator defines an infinite looping with the intention to waste the computation and storage resources within the system.}
\subsection{Software Guard Extensions (SGX)}
\wjss{We use Intel's SGX~\cite{hoekstra2013using} as a TEE, following the wide adoption~\cite{mckeen2016intel, tramer2018slalom, tople2018privado, hunt2018chiron}.
%
SGX provides the extended CPU instruction set to initialize a secure address space, so-called \emph{enclave}.
Applications executed in an enclave can resist eavesdropping and tampering from the outside world.
More precisely, the enclave is isolated from the operating system, hypervisor, and even other enclaves created in the same host.}
%

\wjss{SGX supports \emph{remote attestation} mechanism which enables (1) proving to a remote client the authenticity of an enclave, and (2) creating an encrypted and authenticated communication channel between the enclave and the remote client.
In addition to the identity of the created enclave, the remote client can verify that the enclave is not tampered with and the enclave is running on a genuine platform.
Concretely, the enclave signs the contents by using a group signature scheme and generates a proof known as \emph{attestation}.
The attestation then can be verified by the remote client by resorting to Intel Attestation Service (IAS).}
%

\wjss{However, Intel's SGX is susceptible to well-known side-channel attacks~\cite{xu2015controlled, shinde2015preventing} and rollback attacks~\cite{kaptchuk2019giving}, which can destroy its properties of confidentiality and integrity, respectively.
In this paper, we are fully aware of the two kinds of attacks and our concerns can be seen in Section~\ref{subsec:assumption}.}

\wjss{Additionally, Intel's SGX has a restricted $128$MB Processor Reserved Memory (PRM), but it supports \textit{paging} on Linux system which enables a program inside the TEE to use more memory outside PRM while still preserving both confidentiality and integrity via extra symmetric key cryptography.
Although the memory limitation is removed in SGX 2.0, SGX 2.0 has not been widely used.}

\section{OVERVIEW}
In this section, we present an overview in terms of the system model, threat assumptions and security goals.
\subsection{System Model}\label{system}
\begin{figure}
    \centering
    \includegraphics[width = \columnwidth]{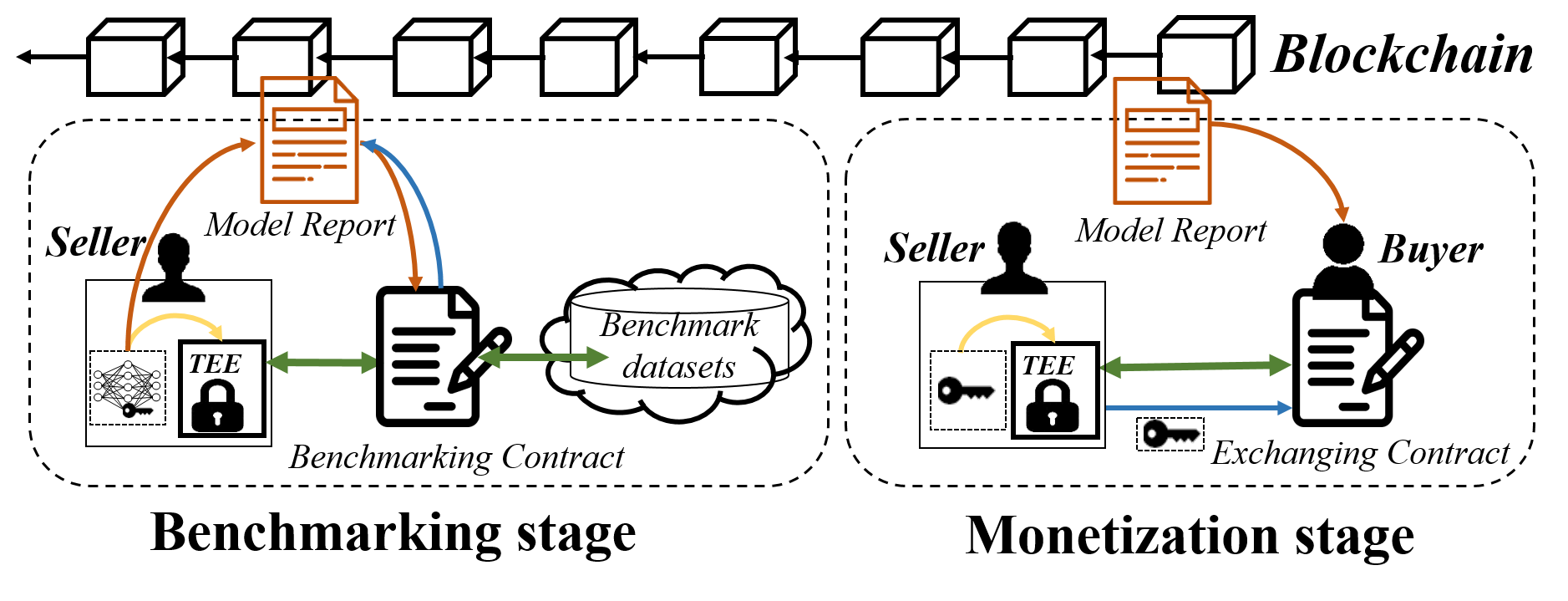}
    \caption{Overview of model marketplace.}
    \label{fig:system}
    \vspace{-15pt}
\end{figure}
At a high level, our proposed decentralized marketplace consists of four major entities, \emph{i.e.}, \emph{model seller}, \emph{model buyer}, \emph{blockchain}, and \emph{external benchmark datasets} shown in Fig.~\ref{fig:system}.
Model seller owns trained models and wants to monetize them, while model buyer is the entity who aims to purchase some trained models pertaining his interests.
Meanwhile, they want to achieve such a model-money swapping process in a decentralized setting, where there is no trusted third party enforcing the whole swapping process.

But without the trusted third party, how to guarantee fairness for the sellers and buyers becomes challenging.
First, since the sellers and buyers in our system are individuals that have no enforced regulations, both of them can cheat in the swapping process to maliciously gain extra benefits.
Second, for sellers, they also demand a transparent pricing mechanism to avoid destructive price competitions in the marketplace~\cite{benhamouda2019supporting}.

To address the two requirements above and enable a healthy ecosystem for the marketplace, we construct the swapping process on the blockchain as a starting point, and further use model performance (\emph{e.g.}, accuracy and robustness) as the important factors for deciding the model price. 
Specifically, before a model can be sold, it is involved in a benchmarking stage where its performance will be evaluated to decide the price used in the later monetization stage.
Here, for ensuring the integrity of the benchmarking process, we follow a trending practice~\cite{zhang2016town} to securely relay authenticated benchmark samples from the benchmarking datasets outside the blockchain, and resort to a TEE for its correctness-enforced executions.
Both the model and the relayed benchmark samples will be put inside the TEE to execute the benchmarking with integrity guarantees, and finally the benchmarked performance will be anchored on the blockchain to guide the later model pricing.

After the benchmarking stage, the model is ready to be sold, and buyers can readily browse the marketplace and choose the model of their interests.
The model monetization stage between a buyer and a seller begins when a buyer deposits money on the blockchain, and finalizes by finishing a fair model-money swapping process which ensures that the seller gets the money as rewards \emph{iff} the buyer can later correctly retrieve the model of his/her interest.

\subsection{Threat Assumptions}\label{subsec:assumption}
\noindent\textbf{Seller and buyer.} We consider that the model seller and the model buyer are malicious, given that they are both loosely regulated.
Specifically, the model sellers might want to forge the benchmark results for selling the model at a higher price, and they might also want to reap the money rewards  without delivering the model or just delivering an incorrect model (\emph{i.e.}, which is inconsistent with the benchmarked one) to the buyer.
While for the model buyers, they might want to repudiate the payment after obtaining the model that they are intended to purchase.

\noindent\textbf{TEE.} We assume that each seller's computer is equipped with an SGX~\cite{mckeen2016intel}, frequently called as enclave.
But instead of readily assuming the SGX as a totally protected environment which perfectly guarantees confidentiality and integrity of the executions, we carefully take into account the recent side-channel attacks \cite{xu2015controlled, shinde2015preventing} and rollback attacks~\cite{kaptchuk2019giving}.
Particularly, the goal of side-channel attacks is to reveal the secret transmitted into the enclave via observing side-channel information during the enclave runtime, \emph{e.g.}, cache timing and power consumptions.
We are aware that although many efforts have been devoted to concealing side-channel leakages~\cite{ahmad2018obliviate, rane2015raccoon, costan2016intel} inside the enclave, they cannot address physical access-based attacks and still lack systematic protections that can perfectly address all side-channel leakages.
\emph{\textbf{Therefore, in this paper we follow the TEE security assumption in~\cite{tramer2017sealed} and consider that the enclave only ensures the confidentiality of the attestation keys rather than sealing keys.}}
In contrast, rollback attacks target damaging the enclave's execution integrity by replaying old states to the enclave.
Here, we consider that a malicious host of the SGX (\emph{e.g.}, the seller) might compromise the benchmark results by initiating rollback attacks.

In addition, we consider that the host may launch multiple enclaves at the same time, but they can be identified via a unique fresh random nonce, and meanwhile, we assume that the probability of two co-existing enclaves with the same nonce is negligible~\cite{tramer2017sealed}.

\noindent\textbf{Benchmark datasets.}
\wjss{We assume that the benchmark datasets are correctly constructed and unrevealed to model sellers before the benchmarking stage.
The assumption is consistent with the methodology of the existing works~\cite{hendrycks2019benchmarking, hendrycks2019augmix, ovadia2019can}, that is, the models to be benchmarked are assumed not trained on the benchmark datasets.}

\wjss{We argue that the assumption is feasible in the real world.
For example, our marketplace can make a commercial contract with the designers of the benchmark datasets.
%
More practically, we can connect our marketplace with Kaggle~\cite{Kaggle}, a currently popular data science competition platform.
In Kaggle, a task organizer is able to recruit a group of competitors to build ML models and after collecting the competitors' models, he has the ability to reveal a test dataset and uses it to evaluate the competitors' models.
Here, we assume that the task organizer is honest and his test datasets can be stored in a system which can be authentically accessed, \emph{e.g.,} via an HTTPS-enabled website.}

\wjss{Besides, we assume that a subset of data used to benchmark a model is generated by randomly sampling from the whole benchmark dataset.
It means that the subset of data has the representative distribution of the whole benchmark dataset~\cite{yang2018lightweight, gretton2012kernel, anderson1994two}}, and thus different models are fairly benchmarked with different test samples.

%
%
\noindent\textbf{Remark.} We additionally trust the blockchain for integrity and availability, but not for privacy, following a widely adopted threat model of the blockchain~\cite{kosba2016hawk}.

\vspace{-0.5em}
\subsection{Security Goals}\label{sec:goals}
In this paper, we want to build a secure and decentralized model marketplace that implements correct model benchmarking and fair model monetization through the blockchain.
In particular, we have the following three security goals:


\noindent\textbf{Model privacy.} The seller's model is never revealed to anyone other than the buyer who has bought the model.

\noindent\textbf{Model correctness.} The model for selling is correctly benchmarked using the external benchmark datasets, and the buyers is able to obtain the correct model which exactly matches the model's report  browsed by the buyer.

\noindent\textbf{Exchanging fairness.} The buyer should obtain the model once his money is paid to the seller, and the seller should get paid once the model is revealed to the buyer.

\subsection{Real-world Application Scenario}\label{sec:scenario}
\wjss{Consider a real-world application scenario in an ML competition platform, such as Kaggle~\cite{Kaggle}.
Our design can be applied into this real-world application scenario to achieve some desirable security properties.}

\wjss{In the scenario, a competition organizer has both the public training set and private test set and solicits multiple model developers for solving his ML task.
To start with, the competition organizer can publish an ML task along with a public training set and an unrevealed test set.
Then, parties of interest compete for the task by training ML models on the public training set and submit their trained models.
Lastly, to choose the best model from the submitted models with varying quality, the organizer can launch a testing process to benchmark the performance of the submitted models by using the test set that those models never see before.}

\wjss{Yet, it might not be easy to conduct the testing process between the competition organizer and participating parties who are mutually distrustful.
The reasons are three-fold.
Firstly, due to intellectual property issues, participating parties (also called as model developers) might not expect to disclose their trained models to the organizer before being paid, since they spend a lot of efforts, e.g., time and computation, to train their models.
Secondly, the organizer needs to conduct a correct testing process, so as to obtain the authentic performance of the models and choose the best one.
Lastly, the organizer gets the model of his interest \emph{iff} the model's owner receives the money paid by the organizer.}

\wjss{In light of the above concerns, the organizer and participating parties can leverage our design involving the benchmarking and monetization stages to achieve two functionalities: \emph{correct model testing} and \emph{fair model-money swapping}.
In such a setting, the benchmark data sets used in the benchmarking stage are provided by the organizer, which are revealed and authentically accessed in the testing process.
Also, the benchmark data sets are trusted by the participating parties for the current ML task.
For further enforcing the authenticity, we can require the organizer to commit to his benchmark data sets on the blockchain before publishing his ML task; the corresponding commitments enable checking the consistency of the benchmark data sets when the organizer reveals them during the testing process.}

\section{CONCRETE DESIGN}
In this section, we present the detailed designs of the benchmarking stage and the monetization stage. We particularly pay the major attention to the benchmarking stage, since it is a crucial step serving for the monetization stage.
\subsection{Benchmarking Stage}\label{sec:roadmap_bench}
\begin{figure}
    \centering
    \includegraphics[width=0.9\columnwidth]{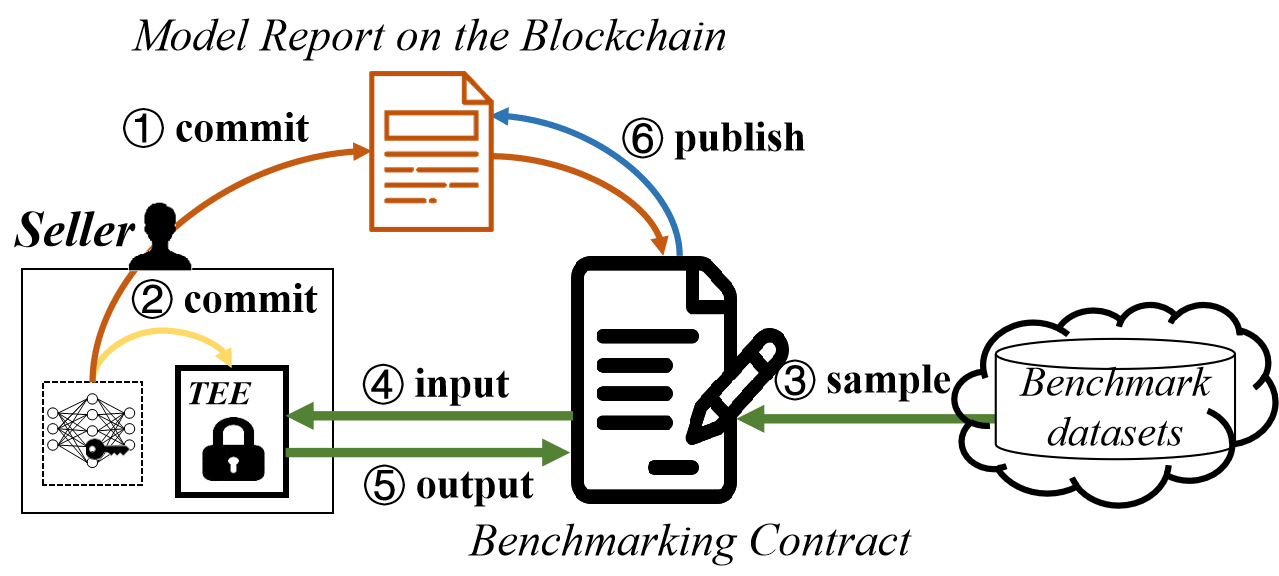}
    \caption{Workflow of the benchmarking stage.}
    \label{fig:benchmarking}
    \vspace{-15pt}
\end{figure}
\subsubsection{\wjss{Intuition}}
\wjss{Ensuring exchange fairness is a crucial requirement for our marketplace.
Specifically, exchange fairness contains two-level meanings.
First, a model-money swapping process is fairly completed, where a seller receives money from a buyer \emph{iff} the buyer obtains the model.
This level of meaning is easily understood from literature~\cite{lu2018zebralancer, li2019toward}.
Besides the fairness during swapping, the buyer should indeed get an authentic model, according to a previously agreed price with respect to the model's authentic performance, which is the second-level meaning of fairness.
The benchmarking stage exactly aims to correctly evaluate models and obtain their authentic performance.
To obtain the model's performance, it is a common practice to benchmark the model over representative benchmark data.
Furthermore, to guarantee the authenticity of model performance while respecting the model privacy, we resort to the TEE residing on a seller's local device.
In addition to the execution correctness, we need to transmit trustworthy benchmark data into the TEE, so as to further enforce the correctness of benchmarking.
After benchmarking, each seller's model is given a transparent price on the smart contract, according to the benchmarked performance.
In this stage, the TEE's remote attestation mechanism enables any third party to publicly verify the correctness of benchmarking.
Meanwhile, the blockchain plays the role of keeping the consistency of off-chain and on-chain data as well as offering a public interface for buyers to purchase the models of their interest.}
\subsubsection{Design Workflow and Concerns}\label{subsec:concerns}
Suppose that a seller possesses a trained model and registers \wjs{with} our marketplace to sell it.
The workflow of model benchmarking is as follows:
\textsf{\circled{1}} The seller \wjs{commits to} his model as $Com_m$ on the blockchain via publishing a model report.
%
This event of publishing invokes the execution of the Benchmarking Contract (called as \textsf{BM-Contract} hereafter).
The \textsf{BM-Contract} reads $Com_m$ from the blockchain.
\textsf{\circled{2}} The seller \wjs{commits to} and loads his model in a local TEE.
Here, we do not distinguish the role of the seller and the TEE's host.
\textsf{\circled{3}} The \textsf{BM-Contract} requests and in response receives \wjss{a set of test data (referred to `samples' hereafter) randomly sampling from the benchmark dataset.}
\textsf{\circled{4}} The seller gets the samples and $Com_m$ from the \textsf{BM-Contract} and relays them into the TEE.
The TEE checks whether or not the loading model is indeed the one committed by $Com_m$ before.
If the checking result is true, it proceeds to run the model evaluation on the samples; otherwise, it aborts.
\textsf{\circled{5}} The TEE returns a proof containing the evaluation outputs \wjs{(i.e., model $m$'s performance)} to the \textsf{BM-Contract}.
\textsf{\circled{6}} If the proof is verified to be true, the \textsf{BM-Contract} gives a price based on the evaluation outputs and then publishes the evaluation outputs and price on the blockchain.
\wjss{It is noteworthy that the correctness of benchmarking indicates two points: (1) the benchmarking process is correctly executed and returns authentic outputs; (2) the used benchmark data are authentically relayed without bias.}
To securely design the above workflow, we have the following three concerns:

\noindent(a) \emph{How to guarantee that the model loaded inside the TEE is the off-chain stored model?}
Our intuition is to ensure that the model committed inside the TEE is identical to the one committed on the blockchain, and meanwhile, the on-chain commitment to the off-chain stored model should be authentic.
With the guarantees, this concern can be canceled.

\vspace{2pt}
\noindent(b) \emph{How to ensure that the benchmark samples directly imported into the TEE are authenticated while reducing on-chain storage overhead?}
This concern is not intuitive in the workflow and comes from the following two practical issues.
First of all, the benchmark samples received by the \textsf{BM-Contract} might not be trustworthy, suppose that the samples now are directly stored on the contract.
Specifically, although the benchmark samples can be from an authenticated source (\emph{e.g.}, an HTTPS-enabled website), smart contract does not support network access and needs a message repeater, causing the lack of an end-to-end authentication for the transmitted samples~\cite{Greenspan, zhang2016town}.
%
Secondly, the benchmark samples can be large, and thus the samples should not be directly stored on the smart contract, for practicality.

\vspace{2pt}
\noindent(c) \emph{How to convince that the execution of benchmarking inside the TEE is indeed correct considering the TEE vulnerable to rollback attacks?}
This concern is caused by the TEE's fundamental limitations that the TEE relies on the host for storing its intermediate states during the execution.
%
The malicious host may relay stale states into the TEE, which can lead to the incorrect execution of the TEE, even if the TEE itself is correct~\cite{matetic2017rote, kaptchuk2019giving}.

For ease of presentation, \wjs{we give a detailed design mainly taking concern (a) into consideration.} We left the last two concerns in Section~\ref{sec:refinement}.

\begin{table}[htbp]
 \centering
 \caption{\label{tab:report} \wjss{Notations used in our design.}}
 \begin{tabular}{ll}
  \toprule
  \textbf{Notation} & \textbf{Description} \\
  \midrule
  $\textcolor[rgb]{0.5,0.1,0.1}{\mathcal{S}}$ & the account address of a seller on the blockchain \\
  $\textcolor[rgb]{0.5,0.1,0.1}{\mathcal{B}}$ & the account address of a buyer on the blockchain \\
  $\textcolor[rgb]{0,0,1}{Addr}_{m}$ &  the storage address of encrypted model $m$   \\
  $\textcolor[rgb]{0,1,0}{Prog}_{m}$ &  the program binary implementing model $m$ \\
  $\textcolor[rgb]{0,1,0}{prog}$ & the program executed inside a seller's enclave during \\
         & the benchmarking stage \\
  $ID_m$ &  the model identifier    \\
  $k_m$ & the symmetry secret used to encrypt model $m$ \\
  $Com_k$ & the commitment to secret $k$   \\
  $Com_m$ & the commitment to model $m$ \\
  $p_k$ & a proof of proving a model indeed encrypted by $k_m$ \\
  $p_{ck}$ & a proof of proving the construction correctness of $Com_k$ \\
  $p_{cm}$  & a proof of proving the construction correctness of $Com_m$ \\
  \textsf{sid} & the session identifier \\
  \textsf{eid} & the enclave identifier \\
  $st_c$ & the state on the smart contract \\
  $T$ & time point on the smart contract \\
  $pk_{\textcolor[rgb]{0.5,0.1,0.1}{\mathcal{B}}}$ &  the public key of a buyer \\
  \textsf{mCE} & the metric describing a model's corruption robustness\\
  \textsf{mFP} & the metric describing a model's perturbation robustness \\
  \textsf{ce} & the metric describing a model's nature accuracy \\
  $price$ & the price of a model \\
  \bottomrule
 \end{tabular}
\end{table}

\subsubsection{Seller Registration}\label{subsec:preparation}
Seller $\textcolor[rgb]{0.5,0.1,0.1}{\mathcal{S}}$ registers with our model marketplace and publishes a model report to the blockchain.
\wjss{The model report contains some public terms which indicate the successful registration of Seller $\textcolor[rgb]{0.5,0.1,0.1}{\mathcal{S}}$.
In addition, the model report will be appended with new terms, such as \textsf{mCE}, \textsf{mFP} and \textsf{ce}, $price$, after the benchmarking stage.}

We proceed to explain the necessary terms on the model report shown in TABLE~\ref{tab:report} as following:
%
$\textcolor[rgb]{0,0,1}{Addr}_{m}$ is the storage address of model $m$; we suppose that seller $\textcolor[rgb]{0.5,0.1,0.1}{\mathcal{S}}$ can encrypt and store his model in a decentralized storage system (\emph{e.g.,} IPFS).
\wjss{$\textcolor[rgb]{0,1,0}{Prog}_{m}$ is the program binary containing the codes corresponding to the network architecture of machine learning model $m$, which does not reveal private model parameters.}
\wjss{$\textcolor[rgb]{0,1,0}{prog}$ not only contains the codes corresponding to the model network architecture, but also contains the codes to calculate the performance metrics and compute a commitment $Com_m^{'}$.}
$ID_m$ is the model identifier which is generated by hashing model $m$'s storage address and the respective seller's account address; suppose that a model owns a unique model report on the blockchain, so a model report can be identified by $ID_m$, and the updated model does not share the same identifier.
\wjss{$k_m$ is the symmetry secret used to encrypt the model.}
\wjss{$p_k$ is the proof which proves that the remotely stored model is indeed encrypted by secret $k_m$ by using a symmetry encryption algorithm.}
$p_{ck}$ and $p_{cm}$ are the proofs used to prove that commitments $Com_k$ and $Com_m$ are authentic, respectively.
%
It is noteworthy that $p_k$, $p_{ck}$ and $p_{cm}$ are asymmetry knowledge proofs, which can be generated by launching additional enclaves~\cite{tramer2017sealed}.
\wjss{We omit the detailed explanation for them.
Other relevant notations in TABLE~\ref{tab:report} which are used in our design will be elaborated in the later section.}

In addition, the model report also contains some useful information for sell, such as the intended use of the model and the model performances (\emph{e.g.}, error rate\footnote{The error rate is estimated by the quantity of samples being  incorrectly  classified  divided  by  the  total  quantity of samples being tested via a confusion matrix.\looseness=-1}, accuracy, confidence level), by following a trending practice.
Here, the model performances are evaluated on a local holdout test by the seller, which \wjs{may be not reliable to buyers} and should be distinguished from the benchmarked performance we introduce below.\looseness=-1

\subsubsection{Performance Metrics of Benchmarking Model}
The publishing of the model report invokes the execution of the \textsf{BM-Contract}, which means entering the benchmark stage.
\wjs{The contact interacts with the seller to evaluate the model performance fed with standard benchmark datasets.}
The seller would agree on the interaction if he wants to promote the sale of his model.
We note that the program logic of the \textsf{BM-Contract} is pre-defined by a group of maintainers in the model marketplace like the blockchain miners.
For simplicity of introducing the model benchmarking, we focus on the setting of \textit{object recognition}. Particularly, models are used to identify the class for the objects in a given image based on its raw pixels (\wjss{other settings will be discussed in Section~\ref{sec:discuss}}).

We benchmark models using three performance metrics, namely, corruption robustness, perturbation robustness and nature accuracy.
\wjss{In addition to the nature accuracy, we stress that the
corruption robustness and perturbation robustness separately measuring the robustness of model against corrupted and perturbed data are crucial for MLaaS, especially in the safety-critical scenarios.}
For example, we do not expect that a traffic sign classifier breaks down due to a sign image added by a crafted perturbation~\cite{carlini2017towards, madry2017towards, papernot2017practical}.
To achieve this, we use the standard benchmark datasets, where data are added with the common corruptions and perturbations.
They include a corruption dataset---ImageNet-C, a perturbation dataset---ImageNet-P~\cite{hendrycks2019benchmarking, hendrycks2019augmix, ovadia2019can}.
Additionally, we use ImageNet validation dataset for estimating the nature accuracy.
The former two datasets are derived from the ImageNet validation dataset~\cite{ILSVRC15} applied with 15 corruptions and 10 perturbations, respectively.
Recall that the benchmark datasets are assumed unrevealed to the model sellers before their models are trained (see Section~\ref{subsec:assumption}).

Specifically, the two robustness metrics include the mean Corruption Error (\textsf{mCE}) and the mean Flip Rate (\textsf{mFP}).
\textsf{mCE} is a mean value of corruption errors on ImageNet-C, reflecting a model's \textit{corruption robustness}.
\textsf{mFP} is a mean value of the perturbation errors on ImageNet-P, measuring a model's \textit{perturbation robustness}.
The standard metric for measuring the nature accuracy of a model is denoted as clean error (\textsf{ce}) on the ImageNet validation dataset without containing corrupted and perturbed data.
We note that 1.0 minus \textsf{ce} that equals to the nature accuracy.
Notably, \textsf{mCE} and \textsf{mFP} can consistently represent a model's robustness reliably, since the two metrics do not have a fundamental trade-off relationship and they can be enhanced together~\cite{hendrycks2019benchmarking, hendrycks2019augmix}.
Additionally, a lower value indicates a stronger robustness for both the metrics.

The aforementioned metrics are calculated inside the enclave.
Concretely, the \textsf{BM-Contract} will require the seller to load the corresponding calculation programs and execute the programs correctly inside the enclave, so that the contract can release itself from the complex computation.
We define those programs as $\textcolor[rgb]{0,1,0}{prog}_{\textsf{mCE}}$, $\textcolor[rgb]{0,1,0}{prog}_{\textsf{mFP}}$ and $\textcolor[rgb]{0,1,0}{prog}_{\textsf{ce}}$, respectively.
They can be wrapped as the programs embedded in program $\textcolor[rgb]{0,1,0}{prog}$.
In such a way, the contract finally obtains the computed statistics, \emph{i.e.}, \textsf{mCE}, \textsf{mFP} and \textsf{ce}, \wjs{over the corresponding benchmark data from the enclave.}
As demonstrated above, estimating the three metrics needs different benchmark datasets, ImageNet-C, ImageNet-P and ImageNet validation dataset, respectively.
\wjss{It implies that there occurs a multi-step interaction between the host and the enclave, since the host needs to relay the requested samples from different benchmark datasets into the enclave for running the process of model evaluation.}
%
%
In addition, the size of the benchmark samples apparently is large, so it is impractical to store the samples on the blockchain.
We optimize this issue in Section~\ref{sec:refinement}.\looseness=-1

\begin{figure}[t!]
\fbox{
\begin{minipage}[t]{0.94\linewidth}
\footnotesize

\hspace{-3pt}\textsf{BM-Contract} (\textcolor[rgb]{0.5,0.1,0.1}{$\mathcal{S}$}, \textcolor[rgb]{0,0,1}{${Addr}_{m}$}, $ID_m$, \textcolor[rgb]{0,1,0}{$prog$}, $Com_m, T_1, T_2, T_3$)
\vspace{3pt}

\scriptsize
\hspace{-3pt}$1$
\hspace{5pt}
$\mathbf{Trigger}$: Upon receiving the publishing event of a model report:

\hspace{-3pt}$2$
\hspace{15pt} assert $ID_m$

\hspace{-3pt}$3$
\hspace{15pt} set $st_c$ = TRIGGERED

\hspace{-3pt}$4$
\hspace{15pt} \textcolor[rgb]{0.1,0.1,0.8}{/* sid is the session id */}

\hspace{-3pt}$5$
\hspace{15pt} send (\textsf{sid}, \textcolor[rgb]{0.5,0.1,0.1}{$\mathcal{S}$}, $``\textsf{install}"$, $\textcolor[rgb]{0,1,0}{prog}$, $T_1$) to \textcolor[rgb]{0.5,0.1,0.1}{$\mathcal{S}$}

\vspace{3pt}
\hspace{-3pt}$6$
\hspace{5pt}$\mathbf{Commit}$: Upon receiving transaction containing \textsf{eid}: 

\hspace{-3pt}$7$
\hspace{15pt} assert $T< T_1$

\hspace{-3pt}$8$
\hspace{15pt} send (\textsf{sid}, \textcolor[rgb]{0.5,0.1,0.1}{$\mathcal{S}$}, $``\textsf{commit}"$, $Com_m$, $T_2$) to \textcolor[rgb]{0.5,0.1,0.1}{$\mathcal{S}$}

\vspace{3pt}
\hspace{-3pt}$9$
\hspace{5pt}
$\mathbf{Request}$: Upon receiving transaction $tx_c^{\textcolor[rgb]{0.5,0.1,0.1}{\mathcal{E}}}$:

\hspace{-6pt}$10$
\hspace{15pt} assert $T< T_2$

\hspace{-6pt}$11$
\hspace{15pt} set $st_c$ = COMMITTED

\hspace{-6pt}$12$
\hspace{15pt} \textcolor[rgb]{0.1,0.1,0.8}{/* sampling randomness will be introduced */}

\hspace{-6pt}$13$
\hspace{15pt} \textcolor[rgb]{0.1,0.1,0.8}{/* in Section~\ref{sec:refinement} */}

\hspace{-6pt}$14$
\hspace{15pt} \textsf{REQUEST}(\textcolor[rgb]{0.5,0.1,0.1}{$\mathcal{S}$}, \textsf{GET}, \textsf{URL}, \textsf{params})

\hspace{-6pt}$15$
\hspace{15pt} wait to receive (samples)

\hspace{-6pt}$16$
\hspace{15pt} \textcolor[rgb]{0.1,0.1,0.8}{/* optimize it in Section~\ref{sec:refinement} */}

\hspace{-6pt}$17$
\hspace{15pt} add (samples) to ledger

\hspace{-6pt}$18$
\hspace{15pt} send (\textsf{sid}, \textcolor[rgb]{0.5,0.1,0.1}{$\mathcal{S}$}, $``\textsf{evaluate}"$, samples, $T_3$) to \textcolor[rgb]{0.5,0.1,0.1}{$\mathcal{S}$}

\vspace{3pt}
\hspace{-6pt}$19$
\hspace{5pt}
$\mathbf{Publish}$: Upon receiving transaction $tx_o^{\textcolor[rgb]{0.5,0.1,0.1}{\mathcal{E}}}$:

\hspace{-6pt}$20$
\hspace{15pt} set $st_c$ = REQUESTED

\hspace{-6pt}$21$
\hspace{15pt} assert $T< T_3$

\hspace{-6pt}$22$
\hspace{15pt} add ($ID_m$, \textsf{outp}) to ledger
\end{minipage}
}
\caption{\wjss{Definition of the Benchmarking Contact.}}\label{fig:tcontract}
\vspace{-15pt}
\end{figure}

\subsubsection{Model Benchmarking with Correctness Guarantees}
We realize the workflow of model benchmarking with a protocol $\mathbf{Prot}_{BM}$ by utilizing a trusted hardware-based zero-knowledge proof (\emph{i.e.}, sealed-glass proofs, SGPs)~\cite{tramer2017sealed} and smart contract in the blockchain setting.
In this protocol, the seller is a \emph{prover} while the \textsf{BM-Contract} plays a role of a \emph{verifier}.
The seller needs to prove that the process of model benchmarking is correct, which implies (1) the benchmarking process is executed with integrity guarantees; (2) the model is benchmarked on the authentically relayed benchmark samples.
\wjss{Note that the trusted hardware is not expected to provide the confidentiality, since the seller's model does not leave its local device.}

First of all, we leverage the functionalities of the SGPs, $\mathcal{F}_{SGP}$, enabling the seller to convince the \textsf{BM-Contract} that he correctly evaluates his model on \wjs{the benchmark samples which are fetched from an authenticated source outside the blockchain} and returns the evaluation results.
We highlight two points when leveraging the SGPs:
(1) the seller should firstly commit to his model in the enclave before receiving the test samples.
(2) the enclave can bind its public key $pk_{\textsf{TEE}}$ to an account \textcolor[rgb]{0.5,0.1,0.1}{$\mathcal{E}$} on the blockchain, and meanwhile, the respective private key of $pk_{\textsf{TEE}}$ is protected to generate attestations (\emph{e.g.} Intel's EPID signatures).
\wjss{Concretely, the first point aims to prevent the seller from faking his model by observing the verifier's inputs.}
\wjss{The second point enables the attestations of the seller's enclave to be regarded as the blockchain transactions and to be verified via the transaction verification protocol of the underlying blockchain, thereby eliminating the on-chain cost of verifying the attestation on the \textsf{BM-Contract}~\cite{zhang2016town}.}

\wjss{Secondly, we resort to smart contract and blockchain for enforcing the execution correctness of the workflow.}
Here, we emphasize a vital feature, that is, a trusted clock provided by the underlying blockchain.
\wjss{As highlighted in the previous work~\cite{kosba2016hawk, pass2017formal}, a trusted clock is crucial to achieve financial fairness and computational fairness.}
Although a TEE also owns a clock source, we do not rely on it due to its timer failures~\cite{cheng2019ekiden}.
We define a trusted clock $T$ in a smart contract, and any party can read it.
The clock is incremented when a new block is created, meaning a new round starts, and the smart contract is executed in rounds.
We also preserve other generic features on the blockchain, such as authenticated messages, message batches and money (see~\cite{kosba2016hawk}).\looseness=-1

\begin{figure}[t!]
\fbox{
\begin{minipage}[t]{0.94\linewidth}
\footnotesize

\textsf{Program for the enclave \textsf{eid}}
\vspace{1pt}
\scriptsize

\hspace{-3pt}$1$
\hspace{5pt}$\mathbf{Init}$: ($pk_{\textsf{TEE}}$, $sk_{\textsf{TEE}}$):=$\Sigma_{\cdot}$\textsf{KGen}$(1^{\lambda})$

\vspace{3pt}
\hspace{-3pt}$2$
\hspace{5pt}$\mathbf{Intall}$: On input((\textsf{sid}, \textcolor[rgb]{0.5,0.1,0.1}{$\mathcal{S}$}, $``\textsf{install}"$, \textsf{eid}, \textcolor[rgb]{0,1,0}{$prog$}), $r$, $mem$)

\hspace{-3pt}$3$
\hspace{15pt} parse $r$ as a nonce $N$ of length $\lambda$

\hspace{-3pt}$4$
\hspace{15pt} store (\textsf{eid}, \textcolor[rgb]{0.5,0.1,0.1}{$\mathcal{S}$}, \textcolor[rgb]{0,1,0}{$prog$}, \underline{~~}, \underline{~~}, \underline{~~}) if no (\textsf{eid}, \textcolor[rgb]{0.5,0.1,0.1}{$\mathcal{S}$}, \underline{~~}, \underline{~~}, \underline{~~}, \underline{~~}) is stored

\hspace{-3pt}$5$
\hspace{15pt} return (\textsf{sid}, \textsf{eid}, $N$, $``\textsf{okay}"$)

\vspace{3pt}
\hspace{-3pt}$6$
\hspace{5pt}\wjs{$\mathbf{Resume}$}: On input(\wjs{\textsf{sid}, $``\textsf{resume}"$}, (\textcolor[rgb]{0.5,0.1,0.1}{$\mathcal{S}$}, \wjs{$``\textsf{commit}"$}, \textsf{eid}, $Com_m$, $model$),

\hspace{5pt} $r$, $mem$)

\hspace{-3pt}$7$
\hspace{15pt} store (\textsf{eid}, \textcolor[rgb]{0.5,0.1,0.1}{$\mathcal{S}$}, \textcolor[rgb]{0.5,0.1,0.1}{$prog$}, $model$, \underline{~~}, \underline{~~})

\hspace{-3pt}$8$
\hspace{15pt} \textcolor[rgb]{0.1,0.1,0.8}{/* check $Com_m == Com_m'$ */}

\vspace{3pt}
\hspace{-3pt}$9$
\hspace{15pt} $\sigma_c:=\Sigma_{\cdot}\textsf{Sig}(sk_{\textsf{TEE}}$, \textcolor[rgb]{0,1,0}{$prog$}, $Com_m'$)

\vspace{3pt}
\hspace{-6pt}$10$
\hspace{15pt} return ((\textsf{sid}, \textsf{eid}, $N$, $``\textsf{receipt}"$), $\sigma_c$)

\vspace{3pt}
\hspace{-6pt}$11$
\hspace{5pt}\wjs{$\mathbf{Resume}$}: On input(\wjs{\textsf{sid}, $``\textsf{resume}"$}, (\wjs{ \textcolor[rgb]{0.5,0.1,0.1}{$\mathcal{S}$}, $``\textsf{evaluate}"$}, \textsf{eid}, samples), $r$,

\hspace{5pt}  $mem$)

\hspace{-6pt}$12$
\hspace{15pt} store (\textsf{eid}, \textcolor[rgb]{0.5,0.1,0.1}{$\mathcal{S}$}, \textcolor[rgb]{0,1,0}{$prog$}, $model$, samples, \underline{~~})

\hspace{-6pt}$13$
\hspace{15pt} \textsf{outp}:= \textcolor[rgb]{0,1,0}{$prog$}($model$, samples)

\vspace{3pt}
\hspace{-6pt}$14$
\hspace{15pt} store (\textsf{eid}, \textcolor[rgb]{0.5,0.1,0.1}{$\mathcal{S}$}, \textcolor[rgb]{0,1,0}{$prog$}, $model$, samples, \textsf{outp})

\vspace{3pt}
\hspace{-6pt}$15$
\hspace{15pt} $\sigma_o:=\Sigma_{\cdot}\textsf{Sig}(sk_{\textsf{TEE}}$, \textcolor[rgb]{0,1,0}{$prog$}, \textsf{outp})

\vspace{3pt}
\hspace{-6pt}$16$
\hspace{15pt} return ((\textsf{sid}, \textsf{eid}, $N$, $``\textsf{output}"$, \textsf{outp}), $\sigma_o$)

\end{minipage}
}
\caption{\wjss{Definition of execution programs in the enclave \textsf{eid}.}}\label{fig:program}
\vspace{-15pt}
\end{figure}

\begin{figure}[t!]
\fbox{
\begin{minipage}[t]{0.94\linewidth}
\scriptsize

$\mathbf{Prot}_{BM}$[\textcolor[rgb]{0,1,0}{$prog$}, \textcolor[rgb]{0.5,0.1,0.1}{$\mathcal{S}$}, \textsf{BM-Contract}]
\vspace{1pt}

\textsf{Seller} \textcolor[rgb]{0.5,0.1,0.1}{$\mathcal{S}$}:

\vspace{3pt}
\hspace{-3pt}$1$
\hspace{5pt} On receive (\textsf{sid}, $``\textsf{install}"$, \textcolor[rgb]{0,1,0}{$prog$}, $T_1$) from $\mathcal{G}(\textsf{BM-Contract})$

\hspace{-3pt}$2$
\hspace{15pt} \textcolor[rgb]{0.1,0.1,0.8}{/* the enclave has been initiated and returns identifier \textsf{eid} */}

\hspace{-3pt}$3$
\hspace{15pt} send (\textcolor[rgb]{0.5,0.1,0.1}{$\mathcal{S}$}, \textsf{sid}, $``\textsf{install}"$, \textsf{eid}, \textcolor[rgb]{0,1,0}{$prog$}, $T_1$) to the enclave

\hspace{-3pt}$4$
\hspace{15pt} wait to receive (\textsf{sid}, \textsf{eid}, $``\textsf{okay}"$)

\hspace{-3pt}$5$
\hspace{5pt} On receive (\textsf{sid}, $``\textsf{commit}"$,  $Com_m$, $T_2$) from $\mathcal{G}(\textsf{BM-Contract})$

\hspace{-3pt}$6$
\hspace{15pt} send (\textcolor[rgb]{0.5,0.1,0.1}{$\mathcal{S}$}, \textsf{sid}, \wjs{$``\textsf{resume}"$, ($``\textsf{commit}"$, \textsf{eid}, $Com_m$, $model$, $T_2$)})

\hspace{18pt} to the enclave

\hspace{-3pt}$7$
\hspace{15pt} wait to receive (\textsf{sid}, \textsf{eid}, $``\textsf{receipt}"$, $tx_c^{\textcolor[rgb]{0.5,0.1,0.1}{\mathcal{E}}}$)

\hspace{-3pt}$8$
\hspace{15pt} \textcolor[rgb]{0.1,0.1,0.8}{/* forward $\sigma_c$ as transaction $tx_c^{\mathcal{E}}$ to the blockchain */}

\vspace{3pt}
\hspace{-3pt}$9$
\hspace{5pt} On receive (\textsf{sid}, $``\textsf{evaluate}"$, samples, $T_3$) from $\mathcal{G}(\textsf{BM-Contract})$

\hspace{-6pt}$10$
\hspace{15pt} send (\textcolor[rgb]{0.5,0.1,0.1}{$\mathcal{S}$}, \textsf{sid}, \wjs{$``\textsf{resume}"$, ($``\textsf{evaluate}"$, \textsf{eid}, samples, $T_3$)}) to the enclave

\hspace{-6pt}$11$
\hspace{15pt} wait to receive (\textsf{sid}, \textsf{eid}, $``\textsf{output}"$, \textsf{outp}, $tx_o^{\textcolor[rgb]{0.5,0.1,0.1}{\mathcal{E}}}$)

\hspace{-6pt}$12$
\hspace{15pt} \textcolor[rgb]{0.1,0.1,0.8}{/* forward $\sigma_o$ transaction $tx_o^{\mathcal{E}}$ to the blockchain */}
\vspace{2pt}
\end{minipage}
}
\caption{\wjss{Definition of protocol $\mathbf{Prot}_{BM}$.}}\label{fig:protocol}
\vspace{-15pt}
\end{figure}

As demonstrated in Fig.~\ref{fig:protocol}, $\mathbf{Prot}_{BM}$ shows that a seller who uses the enclave interacts with the \textsf{BM-Contract}.
\wjss{Formally, we define in Fig.~\ref{fig:program} the execution program inside the enclave by combining with the transparent TEE \emph{ideal functionality} which is formalized by the SGPs protocol~\cite{tramer2017sealed}.
We also define the program logic of the \textsf{BM-Contract} demonstrated in Fig.~\ref{fig:tcontract} by using the blockchain functionality which is defined under the UC framework~\cite{kosba2016hawk}.}
For simplicity, we add a wrapper $\mathcal{G}$ on the \textsf{BM-Contract}, representing messages read from the \textsf{BM-Contract}. The interactive procedures are as follows:

\vspace{2pt}
\noindent(a) Seller \textcolor[rgb]{0.5,0.1,0.1}{$\mathcal{S}$} receives the message to install \wjs{programs \textcolor[rgb]{0,1,0}{$prog$}} with deadline $T_1$.
With the input messages from the seller, the enclave then installs \textcolor[rgb]{0,1,0}{$prog$} and returns its identifier \textsf{eid}.

\noindent(b) Seller \textcolor[rgb]{0.5,0.1,0.1}{$\mathcal{S}$} commits to his model before $T_2$.
\textcolor[rgb]{0.5,0.1,0.1}{$\mathcal{S}$} then loads the model into the enclave memory.
The enclave receives $Com_m$ and checks whether or not it is identical to $Com_m^{'}$, which is generated later inside the enclave, and outputs $\sigma_c$.
Similarly, seller sends $\sigma_c$ as transaction $tx_c^{\textcolor[rgb]{0.5,0.1,0.1}{\mathcal{E}}}$.
%
%
\wjs{If the $tx_c^{\textcolor[rgb]{0.5,0.1,0.1}{\mathcal{E}}}$ is true, it sends a request with a \textsf{URL} for sampling the benchmark data.}

\noindent(c) Seller \textcolor[rgb]{0.5,0.1,0.1}{$\mathcal{S}$} gains the benchmark samples and delivers them into the enclave.
The enclave executes the programs and returns the execution output.
\wjs{The execution output finally contains the evaluation results with respective to three performance metrics.}
Attestation transaction $tx_o^{\textcolor[rgb]{0.5,0.1,0.1}{\mathcal{E}}}$ can be verified to assert the correctness of the execution output.

\noindent(d) The \textsf{BM-Contract} \wjs{then determines a price for the model based on the evaluation results.}
Note that \wjs{the evaluation results} and price will be published on the blockchain, associated with the model report.

\subsubsection{Model Pricing}
\begin{table}[htbp]
 \centering
 \caption{\label{tab:pricing} \wjss{Notations used in the model pricing mechanism.}}
 \begin{tabular}{ll}
  \toprule
  \textbf{Notation} & \textbf{Description} \\
  \midrule
  $q_{\textsf{mCE}}$ &  value of the corruption robustness metric \\
  $q_{\textsf{mFP}}$ &  value of the perturbation robustness metric \\
  $C$ &  cost of generating a model \\
  $c$ &  marginal cost of generating a model \\
  $q$ &  model quality \\
  $W$ &  willingness of buying a model \\
  $l_1,l_2$ & marginal willingness of buying a model \\
  $R$ &  seller revenue \\
  $U$ &  buyer utility \\
  $p$ &  model price \\
  \bottomrule
 \end{tabular}
\end{table}
With the obtained model performance, the \textsf{BM-Contract} determines a price for the model in a transparent manner.
Guided by the market rule of high-performance models being highly valuated, we leverage a quality-based pricing method which uses the performance metrics to measure the model quality.
In addition, this pricing method needs to maximize two objectives which concretely are the revenue of sellers and the utility of buyers, for maintaining the long-term running of the model marketplace.

With the concerns above, we define a performance-oriented model pricing mechanism, and this mechanism is to solve a bi-level programming problem~\cite{bard1991some}. Specifically, the up-level problem is to find the optimal solutions to maximize the \textit{revenue} of sellers, and the low-level problem is to find the optimal solutions to maximize the \textit{utility} of buyers. Both the revenue of sellers and the utility of buyers are related to the model performance.

Additionally, we employ \emph{genetic algorithm}~\cite{holland1992adaptation} to solve the defined bi-level programming problem which is a provably NP-hard problem~\cite{bard1991some}. The reason to choose genetic algorithm is that it has been proved solving bi-level problem efficiently~\cite{camacho2015genetic, oduguwa2002bi, calvete2008new, hejazi2002linear}. Then, the solutions guide the \textsf{BM-Contract} to decide a price for a model with given performance. Notice that for ensuring efficiency, the process of solving the bi-level programming problem is conducted in an off-chain TEE component. The solutions finally returned to the \textsf{BM-Contract} are used to guide pricing.

\js{
Note that although the existing model-based pricing (MBP) algorithm~\cite{chen2018model} meets our market rule that better performance deserves better revenue, it may not fit our scenario well with the following reasons.
First, our model marketplace sells models trained on various datasets while the MBP algorithm bids the instances of a model which are trained on the same dataset.
Second, we benchmark selling models with the consistent benchmarking datasets while the MBP algorithm evaluates a model instance with the individual holdout test data.
Most importantly, a well-performed model in a holdout test may have an uncertainty performance in real-world practice~\cite{hendrycks2019benchmarking, hendrycks2019augmix, ovadia2019can}, meaning that the performance based on the holdout test cannot really indicate the model quality.
}

Now, for easy presentation of the bi-level programming problem for our performance-oriented pricing mechanism, we firstly introduce and denote the following necessary factors which also are summarized in TABLE~\ref{tab:pricing}.
\begin{itemize}
\item \textbf{Performance.} \textsf{mCE} and \textsf{mFP} are used to measure the quality of a model, which are the robustness performance metrics trustfully obtained in the model benchmarking stage. We donate them with $q_{\textsf{mCE}}$ and $q_{\textsf{mFP}}$ $\in (0,1]$, respectively.
As for nature accuracy, we suppose that the market would reject the models whose nature accuracy is below a given threshold, e.g., $60.0$\%, since we think that buyers generally are not willing to purchase a model with low nature accuracy, regardless of robustness.
Data pricing depending on the quality of data has been well-studied in the community of data market~\cite{stahl2017name, yu2017data}.
Following these existing work, we use $q_{\textsf{mCE}}$ and $q_{\textsf{mFP}}$ measuring model quality to score another two factors \textit{cost} and \textit{willingness} below.
\item \textbf{Cost.} It describes the expense of generating a model and varies with the model robustness, \emph{i.e.}, spending more costs to get higher model robustness.
It has been widely investigated that, for improving model robustness, model developers usually consume more time and computation to carefully design algorithms and select additional training data~\cite{hendrycks2019augmix, carmon2019unlabeled, stanforth2019labels}.
Specifically, cost $C$ is scored via a linear function of quality, that is, $C=c\cdot\frac{1}{2}(w_1\cdot q_{\textsf{mCE}}+w_2\cdot q_{\textsf{mFP}})$, where $c\in (0,c_{max}]$ is the marginal cost, and $w_1$ and $w_2$ mean the relative weights with the constraint of $w_1+w_2=1$.
The reason of using an increasing linear function instead of a nonlinear one is that \textsf{mCE} and \textsf{mFP} can consistently represent the model quality, guided by~\cite{yu2017data}.
We note that a model quality is $q=\frac{1}{2}(w_1\cdot q_{\textsf{mCE}}+w_2\cdot q_{\textsf{mFP}})$. In addition, $C>0$.~\looseness=-1
\item \textbf{Willingness.}
It describes the preference of paying for a model and also varies with the model robustness.
It is not hard to understand that service providers prefer robust models for providing reliable ML services~\cite{carlini2017towards, madry2017towards, papernot2017practical}.
Specifically, willingness $W$ is scored by an increasing linear function of $W=\frac{1}{2}(l_1\cdot q_{\textsf{mCE}}+l_2\cdot q_{\textsf{mFP}})$, where $l_1,l_2 \sim \textsf{U}(0,1)$ are the marginal willingness for $q_{\textsf{mCE}}$ and $q_{\textsf{mFP}}$, respectively.
In addition, $W>0$.
We treat both $q_{\textsf{mCE}}$ and $q_{\textsf{mFP}}$ equally essential for buyers.
\end{itemize}

With cost and willingness, we define revenue $R$ of seller and utility $U$ of buyer, with respect to a model in quality $q$, when given a take-it-or-leave-it price $p>0$:
$$
R(q,p,x,y) =p\cdot x-C\cdot y
$$
$$
U(q,p,x) =(W-p)\cdot x
$$
Here, $C$ and $W$ are related to $q$. Note that if the given price $p$ is not higher than expense $C$, a seller would not join the market, which indicates Individual Rationality (IR), and thus $p>C$.
We use $x$ to express whether or not to buy the model, and $x=1$ if $W \geq p$, otherwise, $x=0$ if $W<p$. $y$ is used to express whether or not to sell the model.

Then, the \textbf{B}i-\textbf{L}evel programming problem for \textbf{M}odel \textbf{P}ricing mechanism is denoted, named as BLMP, given that the market shows $N$ models with different levels of quality and price, and $M$ consumers:
\begin{itemize}
    \item the up-level problem
    $$\textsf{max}\quad R(q_i,p_i,x_{i,j},y_i)=\sum_{j=1}^M\sum_{i=1}^Np_ix_{i,j}-\sum_{i=1}^NC_iy_i,$$
with constraints $p_i>C_i>0$, and $x_{i,j}=0$ or $1$, $y_i\leq \sum_{j=1}^Mx_{i,j}$ and $y_{i}=0$ or $1$;
    \item the low-level problem
    $$\textsf{max}\quad
    U_j(W_{i,j},p_i,x_{i,j})=\sum_{i=1}^{N}[(W_{i,j}-p_i)\cdot x_{i,j}],$$
with constraints $W_{i,j}\geq p_i>0$, $x_{i_1,j} \cdot x_{i_2,j}=0$ for any $i_1 \neq i_2$ and $x_{i,j}=0$ or $1$.
\end{itemize}

Notice that each of models at most is sold to one consumer.
Solving the bi-level programming problem is not easy, due to the need of solving the low-level problem at each step of an algorithm that finds the solutions for the up-level problem.
Specifically, the decision maker deciding $y_i$ in the up-level should interact with the decision maker deciding $x_{i,j}$ of the low-level to obtain the best solution~\cite{yu2017data, camacho2015genetic}~\looseness=-1.

Lastly, we use genetic algorithm to approximately solve the above problem in finite time steps, following the works~\cite{yu2017data, camacho2015genetic}.
%
%
Genetic algorithm is a stochastic search technique derived from natural selection and natural genetics mechanisms, consisting of the procedures of \textit{initialization}, \textit{fitness evaluation}, \textit{selection}, \textit{crossover} and \textit{mutation}, in which the latter four procedures repeat successively~\cite{holland1992adaptation}.
The algorithm terminates until a previously configured stopping criterion is reached, \emph{e.g.}, 500 iterations used by~\cite{yu2017data}.
It is noteworthy that in the procedure of fitness evaluation, the fitness value must be the revenue value by computing $R(q,p,x,y)$ which is the objective function value in the up-level problem~\cite{camacho2015genetic}.
After obtaining a set of optimal solutions, the \textsf{BM-Contract} is able to construct pricing curves, which is used to determine a price for a model with known performances (such as \textsf{mCE}, \textsf{mFP} and \textsf{ce}).

\noindent \textbf{Remark.} Once a model is sold, a state \textsf{`SOLD'} is updated to its model report. Any identical model cannot be sold twice and updated models are treated as new models.
Recall that the on-chain commitments of each model can be used to filter out the model which has been registered with the blockchain before.

\subsubsection{Design Refinement}\label{sec:refinement}
We now are ready to refine our design by reviewing the remaining two concerns mentioned in Section~\ref{subsec:concerns}.
Recall that concern (b) is how to ensure that the benchmark samples directly imported into the enclave are authenticated \wjs{without the need to store them} on chain;
concern (c) is how to convince that the model evaluation inside the enclave is indeed correct when the enclave is vulnerable to rollback attacks.
\wjss{In this section, we are ready to refine our design by resolving the two concerns with the off-the-shelf techniques, namely, Town Crier~\cite{zhang2016town} and the Enclave-Ledger Interaction (ELI) protocol~\cite{kaptchuk2019giving}.}

\js{
First of all, we need an efficient off-chain intermediator who correctly relays the trustworthy samples into the seller's enclave, and meanwhile leaves the digest of the samples on the blockchain.
Following the design of Town Crier~\cite{zhang2016town}, we can use the ready TEE-based component to realize the intermediator.
In our scenario, an intuition is to let the seller himself create another enclave served as the intermediator.
Despite that it can work, it may be not cost-saving, compared to using a common intermediator with a TEE-enabled platform.
Imagine that multiple sellers' models are benchmarked at the same time with the consistent benchmarking samples.
The common intermediator can request \emph{once} the samples from the HTTPS-enabled website and forward them to the multiple sellers.
However, in the former case, multiple sellers can cause \emph{multiple} requests and leave multiple digests of the same samples on the blockchain, which is comparatively cost-consuming.
}
\noindent\textbf{Solution for Concern (b):} We adopt an alternative approach to randomly choose an existing seller in the marketplace to provide his enclave served as the common intermediator when the \textsf{BM-Contract} is invoked.
Our marketplace can pay fees for the seller's efforts.
We suppose that the chosen seller's enclave $EID$ has bound its public key $pk_i$ to a blockchain account \textcolor[rgb]{0.5,0.1,0.1}{$\mathcal{I}$}.
Notably, we trust the enclave instead of the seller who may have a malicious host.
\wjss{We also employ the enclave to securely generate the randomness for sampling the benchmark dataset.}

Therefore, the refined design is as follows:
The \textsf{BM-Contract} sends a request message to account \textcolor[rgb]{0.5,0.1,0.1}{$\mathcal{I}$}, namely, \textsf{REQUEST}(\textcolor[rgb]{0.5,0.1,0.1}{$\mathcal{S}$}, \textsf{GET}, \textsf{URL}, \textsf{params}, \textcolor[rgb]{0.5,0.1,0.1}{$\mathcal{I}$}) meaning that \textcolor[rgb]{0.5,0.1,0.1}{$\mathcal{I}$} should forward the returned response message to seller \textcolor[rgb]{0.5,0.1,0.1}{$\mathcal{S}$}, after requesting \textsf{URL} with \textsf{params}.
Then, account \textcolor[rgb]{0.5,0.1,0.1}{$\mathcal{I}$} forwards the request message to enclave $EID$.
With the response of the \textsf{URL}, enclave $EID$ generates a digest of the response message, \emph{i.e.}, the benchmark samples, and meanwhile, it signs the digest, which makes the benchmark samples end-to-end authenticated.
The enclave's host then relays the response message (or samples) to \textcolor[rgb]{0.5,0.1,0.1}{$\mathcal{S}$} and sends the signature plus a digest of the samples to the BM-Contract.
Note that this signature can be verified under the blockchain transaction verification protocol.
Last, \textcolor[rgb]{0.5,0.1,0.1}{$\mathcal{S}$} ingests the samples and the signature (by pulling it from the blockchain) into his enclave. The enclave would assert the samples, by re-generating the digest of the samples and verifying the signature.
If the verification result is true, the samples are authenticated, and the enclave continues to execute the remaining program.\looseness=-1

\js{
Secondly, we need a public and tamper-resistant component to enforce
loading the latest execution states of a program into the enclave.
In addition, we desire that it is decentralized, due to our motivation without a TTP.
Thus, we naturally use the blockchain to realize the component via implementing an instance of the Enclave-Ledger Interaction (ELI) protocol~\cite{kaptchuk2019giving}.
The ELI protocol enables the execution states to be encrypted stored on the potentially malicious host, but it considers two design principles for enforcing up-to-date execution states:
(a) Using a counter to associate with the intermediate states at each execution step and the counter can be traced with the help of the blockchain;
%
(b) New input data for executing the program in the next step should be committed to the blockchain in advance.
In addition, the key used to encrypt the states and the randomness for executing the program, at each execution step, are randomly generated by taking the current data of the blockchain as input.
}

\noindent\textbf{Solution for Concern (c):}
%
We recall that the process of model benchmarking can cause the multi-step execution of the enclave.
Suppose that \textcolor[rgb]{0,1,0}{$prog$} is carefully designed to invoke programs $\textcolor[rgb]{0,1,0}{prog}_{\textsf{mCE}}$, $\textcolor[rgb]{0,1,0}{prog}_{\textsf{mFP}}$ and $\textcolor[rgb]{0,1,0}{prog}_{\textsf{ce}}$ with the samples of ImageNet-C, ImageNet-P and ImageNet, respectively.

The instance derived from the existing ELI protocol~\cite{kaptchuk2019giving} is presented with the following constructions.
The enclave $E$ encrypts/decrypts an intermediate execution state $st_i^E$ and program information \textcolor[rgb]{0,1,0}{$prog$} in a step $i$, using a symmetry authenticated encryption scheme ($\textsf{Enc}(\cdot)$, $\textsf{Dec}(\cdot)$) with the deterministic secret $k_i$.
The encrypted states, \emph{i.e.}, $C_{st}\leftarrow \textsf{Enc}(k_i, (st_i^E, \textsf{Hash}(\textcolor[rgb]{0,1,0}{prog})))$, are stored in the host.
Herein, we emphasize two points.
\wjs{First, program information \textcolor[rgb]{0,1,0}{$prog$} includes program \textcolor[rgb]{0,1,0}{${Prog}_{m}$}, the counter of the execution step and programs $\textcolor[rgb]{0,1,0}{prog}_{\textsf{mCE}}$, $\textcolor[rgb]{0,1,0}{prog}_{\textsf{mFP}}$ and $\textcolor[rgb]{0,1,0}{prog}_{\textsf{ce}}$.}
The counter of the execution step is recorded on the blockchain, so as to check the freshness of states stored in the host.
Second, deterministic secret $k_i$ is derived via pseudorandom function $\textsf{F}(\cdot)$ on $sk_{\textsf{TEE}}$ and a hash value $hash_b$ of the blockchain's latest block, namely, $(k_i, r_i)\leftarrow \textsf{F}(sk_{\textsf{TEE}}, hash_b)$.
Random coin $r_i$ is used by the enclave in each execution to avoid the forked execution.
Notably, \wjs{the digest of the samples has been recorded on the blockchain, when the enclave receives authentically relaying benchmark samples.}
The host thus cannot replay the old block with modified samples into the enclave.
As a result, we can ensure that the state data in the enclave at each step is the latest one.

\subsection{Monetization Stage}\label{sec:roadmap_mone}
After the benchmarking stage, buyers can purchase models according to the model performances anchored on the blockchain and then move into the monetization stage.

\subsubsection{\wjss{Intuition}}
\begin{figure}
    \centering
    \includegraphics[width=0.9\columnwidth]{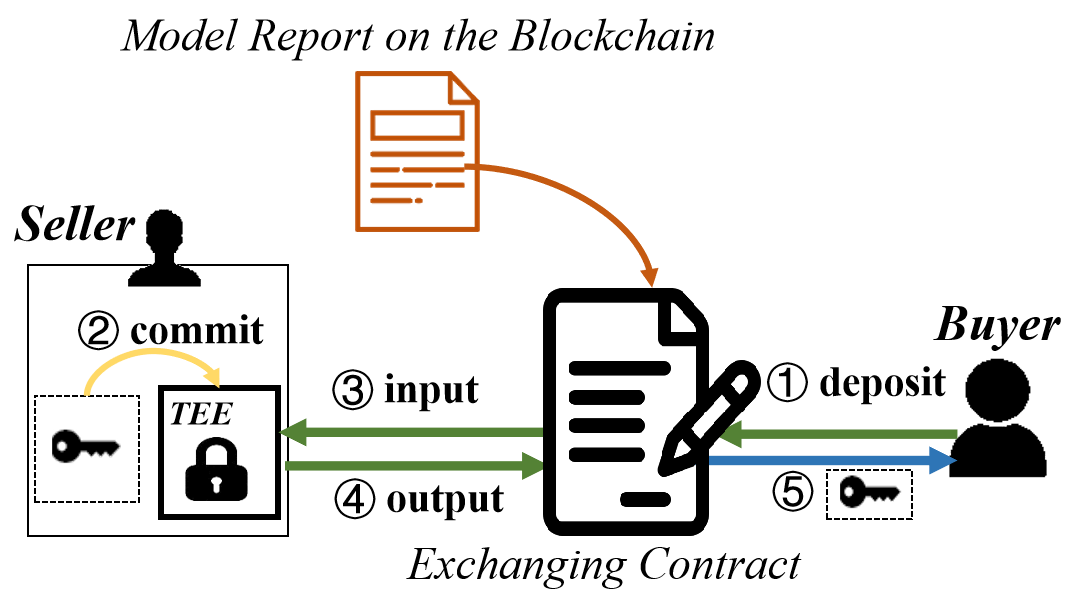}
    \caption{Workflow of the monetization stage.}
    \label{fig:selling}
    \vspace{-15pt}
\end{figure}

\wjss{To truly ensure exchange fairness, enforcing model authenticity via the benchmarking stage is necessary but insufficient, since a fair model-money exchanging procedure has not yet been conducted.
Hence, the monetization stage now is introduced, as shown in Fig.~\ref{fig:selling}.
It aims to achieve exchange fairness between the seller's symmetry secret used to encrypt his model and the buyer's money paid for the model.
Following the technical rule of the benchmarking stage, we continue to leverage the TEE's attestation mechanism and the self-enforcing smart contract to design a fair exchanging protocol.
With the protocol, we fulfill the fairness requirement, that is, the seller receives the money \emph{iff} the buyer obtains the symmetry secret which is used to correctly decrypt the encrypted model.}
\subsubsection{Design Workflow}
Buyers can pull the model reports from the blockchain, and may decide to purchase a model based on his budget and robustness requirements.
The workflow is shown as following:
\textsf{\circled{1}} When deciding to buy, a \textbf{B}uyer can initiate the \textbf{E}xchanging Contract (\textsf{BE-Contract} for short) with a deposit not less than the price of the model he wants to buy and public information including $ID_m$, $p_k$ and $Com_k$.
\textsf{\circled{2}} The model seller \wjs{commits to} and loads his secret (used to encrypt the model) inside the enclave. We note that the seller can encrypt and store his model in a decentralized storage system, \emph{e.g.}, IPFS~\cite{Ben14}, for being retrieved by purchasers later.
\textsf{\circled{3}} The seller then transmits the buyer's public key and the on-chain commitment to the secret into the enclave.
\textsf{\circled{4}} The seller finally returns a secret used to encrypt the model under the buyer's public key.
\textsf{\circled{5}}
The buyer obtains and decrypts the secret using the private key corresponding to $pk_{\textcolor[rgb]{0.5,0.1,0.1}{\mathcal{B}}}$. Meanwhile, the deposit is transferred to the seller's account.
After obtaining the secret, the buyer downloads and decrypts the model.
Note that the buyer should confirm what he obtains is authentic.
This workflow implies a fair exchanging process, in which the component of `commit and reveal' $k_m$ is used to prevent the buyer from being cheated.

\begin{figure}[t!]
\fbox{
\begin{minipage}[t]{0.94\linewidth}

\scriptsize

$\textsf{BE-Contract} (\textcolor[rgb]{0.5,0.1,0.1}{\mathcal{B}}, \textcolor[rgb]{0.5,0.1,0.1}{\mathcal{S}}, pk_{\textcolor[rgb]{0.5,0.1,0.1}{\mathcal{B}}}, ID_m, p_k, Com_k, T_1')$

\vspace{3pt}
\hspace{-3pt}$1$
\hspace{5pt}
$\mathbf{Init}$: Upon receiving a transaction (\textsf{sid'}, \textcolor[rgb]{0.5,0.1,0.1}{$\mathcal{B}$}, $ID_m$, $p_k$, $Com_k$,
\$$Reward_m$):

\hspace{-3pt}$2$
\hspace{15pt} assert $p_k$

\hspace{-3pt}$3$
\hspace{15pt} assert ledger[\textcolor[rgb]{0.5,0.1,0.1}{$\mathcal{B}$}] $\geq$ \$$Reward_m$ $\geq$ $price$

\hspace{-3pt}$4$
\hspace{15pt}\textcolor[rgb]{0.1,0.1,0.8}{/* buyer makes a sufficient deposit */}

\hspace{-3pt}$5$
\hspace{15pt} ledger[\textcolor[rgb]{0.5,0.1,0.1}{$\mathcal{B}$}] $=$ ledger[\textcolor[rgb]{0.5,0.1,0.1}{$\mathcal{B}$}] $-$ \$$Reward_m$


\hspace{-3pt}$6$
\hspace{15pt} set $st_c$ = INITIATED

\vspace{3pt}
\hspace{-3pt}$7$
\hspace{5pt}
$\mathbf{Request}$: Upon receiving a transaction (\textsf{sid'}, \textcolor[rgb]{0.5,0.1,0.1}{$\mathcal{B}$}, $pk_{\textcolor[rgb]{0.5,0.1,0.1}{\mathcal{B}}}$):

\hspace{-3pt}$8$
\hspace{15pt} set $st_c$ = REQUESTED

\hspace{-3pt}$9$
\hspace{15pt} send (\textsf{sid'}, \textcolor[rgb]{0.5,0.1,0.1}{$\mathcal{S}$}, $``\textsf{request}"$, $pk_{\textcolor[rgb]{0.5,0.1,0.1}{\mathcal{B}}}$, $Com_k$) to \textcolor[rgb]{0.5,0.1,0.1}{$\mathcal{S}$}

\vspace{3pt}
\hspace{-6pt}$10$
\hspace{5pt}
$\mathbf{Publish}$: Upon receiving transaction $tx_o'^{\mathcal{E}}$:

\hspace{-6pt}$11$
\hspace{15pt} assert $T \leq T_1'$

\hspace{-6pt}$12$
\hspace{15pt} ledger[\textcolor[rgb]{0.5,0.1,0.1}{$\mathcal{S}$}] $=$ ledger[\textcolor[rgb]{0.5,0.1,0.1}{$\mathcal{S}$}] $\wjs{+}$ \$$Reward_m$

\hspace{-6pt}$13$
\hspace{15pt} set $st_c$ = CLAIMED

\hspace{-6pt}$14$
\hspace{15pt} add ($\textsf{AEnc}_{pk_{\textcolor[rgb]{0.5,0.1,0.1}{\mathcal{B}}}}(k_{m})$, $pk_{\textcolor[rgb]{0.5,0.1,0.1}{\mathcal{B}}}$) to ledger

\vspace{3pt}
\hspace{-6pt}$15$
\hspace{3pt}
\textsf{Time}($T$): assert ($st_c$ = CLAIMED) and ($T > T_1'$)

\hspace{-6pt}$16$
\hspace{15pt} set $st_c$ = ABORTED

\hspace{-6pt}$17$
\hspace{15pt} ledger[\textcolor[rgb]{0.5,0.1,0.1}{$\mathcal{B}$}] $=$ ledger[\textcolor[rgb]{0.5,0.1,0.1}{$\mathcal{B}$}] $+$ \$$Reward_m$
\end{minipage}
}
\caption{\wjss{Definition of the buyer-initiated Exchanging Contact.}}\label{fig:econtract}

\end{figure}

\begin{figure}[t!]
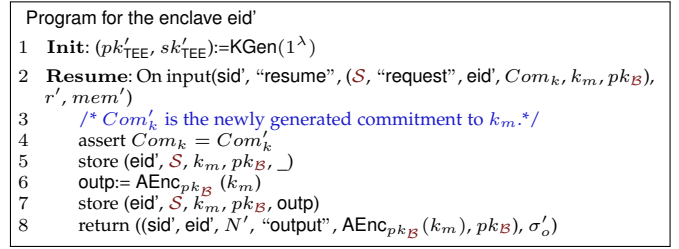

\fbox{
\begin{minipage}[t]{0.94\linewidth}
\scriptsize
\textsf{Program for the enclave \textsf{eid'}}

\vspace{3pt}
\hspace{-3pt}$1$
\hspace{5pt}$\mathbf{Init}$: ($pk_{\textsf{TEE}}'$, $sk_{\textsf{TEE}}'$):=\textsf{KGen}$(1^{\lambda})$

\vspace{3pt}
\hspace{-3pt}$2$
\hspace{5pt}\wjs{$\mathbf{Resume}$}: On input(\textsf{sid'}, $``\textsf{resume}"$, (\textcolor[rgb]{0.5,0.1,0.1}{$\mathcal{S}$},  $``\textsf{request}"$, \textsf{eid'},  $Com_k$, $k_{m}$, $pk_{\textcolor[rgb]{0.5,0.1,0.1}{\mathcal{B}}}$),

\hspace{5pt} $r'$, $mem'$)

\hspace{-3pt}$3$
\hspace{15pt} \textcolor[rgb]{0.1,0.1,0.8}{/* $Com_k'$ is the newly generated commitment to $k_{m}$.*/}

\hspace{-3pt}$4$
\hspace{15pt} assert $Com_k=Com_k'$

\hspace{-3pt}$5$
\hspace{15pt} store (\textsf{eid'}, \textcolor[rgb]{0.5,0.1,0.1}{$\mathcal{S}$}, $k_{m}$, $pk_{\textcolor[rgb]{0.5,0.1,0.1}{\mathcal{B}}}$, \underline{~~})

\hspace{-3pt}$6$
\hspace{15pt} \textsf{outp}:= $\textsf{AEnc}_{pk_{\textcolor[rgb]{0.5,0.1,0.1}{\mathcal{B}}}}$ ($k_{m}$)

\hspace{-3pt}$7$
\hspace{15pt} store (\textsf{eid'}, \textcolor[rgb]{0.5,0.1,0.1}{$\mathcal{S}$}, $k_{m}$, $pk_{\textcolor[rgb]{0.5,0.1,0.1}{\mathcal{B}}}$, \textsf{outp})

\hspace{-3pt}$8$
\hspace{15pt} return ((\textsf{sid'}, \textsf{eid'}, $N'$, $``\textsf{output}"$, $\textsf{AEnc}_{pk_{\textcolor[rgb]{0.5,0.1,0.1}{\mathcal{B}}}}(k_{m})$, $pk_{\textcolor[rgb]{0.5,0.1,0.1}{\mathcal{B}}}$), $\sigma_o'$)

\end{minipage}
}
\caption{\wjss{Definition of execution programs in the enclave \textsf{eid'}.}}\label{fig:programME}
\end{figure}

\begin{figure}[t!]
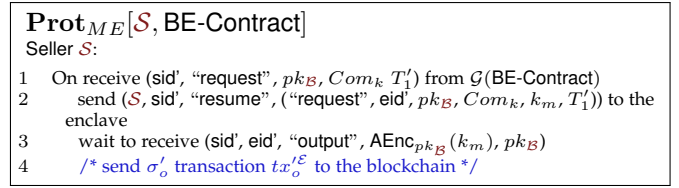

\fbox{
\begin{minipage}[t]{0.94\linewidth}

$\mathbf{Prot}_{ME}[\textcolor[rgb]{0.5,0.1,0.1}{\mathcal{S}}, \textsf{BE-Contract}]$
\vspace{1pt}
\scriptsize

\textsf{Seller} \textcolor[rgb]{0.5,0.1,0.1}{$\mathcal{S}$}:

\vspace{3pt}
\hspace{-3pt}\scriptsize$1$
\hspace{5pt} On receive (\textsf{sid'}, $``\textsf{request}"$, $pk_{\textcolor[rgb]{0.5,0.1,0.1}{\mathcal{B}}}$, $Com_k$ $T_1'$) from $\mathcal{G}(\textsf{BE-Contract})$

\hspace{-3pt}$2$
\hspace{15pt} send (\textcolor[rgb]{0.5,0.1,0.1}{$\mathcal{S}$}, \textsf{sid'}, \wjs{$``\textsf{resume}"$,} ($``\textsf{request}"$, \textsf{eid'}, $pk_{\textcolor[rgb]{0.5,0.1,0.1}{\mathcal{B}}}$, $Com_k$, $k_{m}$, $T_1'$)) to the

\hspace{15pt}enclave

\hspace{-3pt}$3$
\hspace{15pt} wait to receive (\textsf{sid'}, \textsf{eid'}, $``\textsf{output}"$, $\textsf{AEnc}_{pk_{\textcolor[rgb]{0.5,0.1,0.1}{\mathcal{B}}}}(k_{m})$, $pk_{\textcolor[rgb]{0.5,0.1,0.1}{\mathcal{B}}}$)

\hspace{-3pt}$4$
\hspace{15pt} \textcolor[rgb]{0.1,0.1,0.8}{/* send $\sigma_o'$ transaction $tx_o'^{\mathcal{E}}$ to the blockchain */}
\end{minipage}
}
\caption{\wjss{Definition of protocol $\mathbf{Prot}_{ME}$.}}\label{fig:protocolME}
\vspace{-15pt}
\end{figure}

\subsubsection{Model-Money Swapping with Fairness Guarantees}
We now present a detailed design of achieving a fair model-money swapping procedure, where the seller should give an authentic model to the buyer; the model should be consistent to the one the buyer desire to purchase by observing the respective model report on the blockchain.
Meanwhile, the procedure should guarantee the fairness, eliminating the dispute among the buyer and seller.\looseness=-1

We achieve the fair model-money swapping owing to two components: the component of `commit and reveal' and smart contract.
First, the component of `commit and reveal' is used to prove that a seller gives an authentic secret in the presence of not exposing his secret.
We realize it via an SGP protocol which can be seen the enclave program in Fig.~\ref{fig:programME}. Here, we have to emphasize two issues of realizing this component: (a) the seller should commit to the symmetry secret $k_{m}$ for encrypting the model and then deliver it encrypted under buyer \textcolor[rgb]{0.5,0.1,0.1}{$\mathcal{B}$}'s public key $pk_{\textcolor[rgb]{0.5,0.1,0.1}{\mathcal{B}}}$. This way avoids directly delivering the model. (b) the commitment $Com_k'$ to $k_{m}$ inside the seller's enclave should be the same with $Com_k$ on chain. A feasible way is that when entering the market, the seller launches an enclave to generate $Com_k$
and preserves the state of this enclave (refer it to Section~\ref{sec:refinement}); when entering this monetization stage, the seller recovers the preserving state and presents $Com_k'$.
Second, smart contract is utilized to resolve the potential dispute, see the Exchanging Contract in Fig.~\ref{fig:econtract}.\looseness=-1

Now, by integrating SGPs and the \textsf{BE-Contract} initiated by the buyer, the fair model-money swapping is realized with protocol $\mathbf{Prot}_{ME}$.
As illustrated in Fig.~\ref{fig:protocolME}, the seller is required to present that he \wjs{commits to} the symmetry secret $k_{m}$ into the enclave with identifier \textsf{eid'} after the buyer has deposited sufficient money.
Then, if the commitment to $k_{m}$ is successfully checked, the seller sends \textsf{``resume"} message to the enclave ingesting the buyer's $pk_{\textcolor[rgb]{0.5,0.1,0.1}{\mathcal{B}}}$ as input.
The enclave encrypts $k_{m}$ under $pk_{\textcolor[rgb]{0.5,0.1,0.1}{\mathcal{B}}}$ by using an asymmetric encryption scheme \textsf{AEnc} and returns it back.
After successfully verifying $tx_o'^{\mathcal{E}}$, the buyer decrypts the output $\textsf{AEnc}_{pk_{\textcolor[rgb]{0.5,0.1,0.1}{\mathcal{B}}}}(k_{m})$ using the respective private key of $pk_{\textcolor[rgb]{0.5,0.1,0.1}{\mathcal{B}}}$ and obtains $k_{m}$ to decrypt the model stored in the IPFS.

\section{SECURITY ANALYSIS}
With regard to our security goals mentioned in Section~\ref{sec:goals}, we elaborate the security analysis for our presented protocols $\textbf{Prot}_{BM}$ and $\textbf{Prot}_{ME}$.

\subsection{Model Privacy}
In protocol $\textbf{Prot}_{BM}$, a seller's model is locally kept during the process of model benchmarking.
Finally, the seller (or the host) only reveals statistics of the classification error rates on the given samples, without class labels and confidence levels which are used to steal model information~\cite{tramer2016stealing}.
In protocol $\textbf{Prot}_{ME}$, if the process of model swapping is completed as expected, only the buyer can decrypt the encrypted model via a symmetry secret $k_{m}$ which is transmitted under the buyer's public key.
Thus, assuming that the utilized asymmetric encryption algorithm used to encrypt secret $k_{m}$ (i.e., $\textsf{AEnc}_{pk_{\textcolor[rgb]{0.5,0.1,0.1}{\mathcal{B}}}}$ ($k_{m}$)) is semantic secure~\cite{goldwasser1984probabilistic}, anyone excepts that the seller and buyer cannot gain any information about secret $k_{m}$ during the transmission, thereby failing to decrypt the model.
In addition, we assume that published data related to the symmetry key and model on the blockchain do not reveal their values. Specifically, commitments $Com_k$, $Com_m$, $p_{cm}$, $p_{ck}$ and $p_k$ are securely constructed~\cite{goldreich2009foundations}.

\subsection{Model Correctness} Both of protocols guarantee model correctness.
First, protocol $\textbf{Prot}_{BM}$ based on the SGPs protocol~\cite{tramer2017sealed} enables a seller to prove the integrity of the execution and the authenticity of the benchmarked result.
Second, protocol $\textbf{Prot}_{ME}$ uses the SGPs protocol to realize a `commit and reveal' component that a seller commits to a model and then reveals it to a buyer, where the model is the real one this buyer wants to purchase. Note that in the cases of not considering rollback attacks and untrustworthy data feed to \textsf{BM-Contract} (referred to concerns (b) and (c), respectively), model correctness is ensured based on the security of the SGPs protocol. Therefore, we present security analysis for $\textbf{Prot}_{BM}$ in Lemma~\ref{them:theorem_second}, following Tramer~\textit{et al.}'s work~\cite{tramer2017sealed} under the Universal Composability (UC) framework. The case for $\textbf{Prot}_{ME}$ is similar and thus omitted.

\begin{theorem}\label{them:theorem_second}
Assume that the used TEE's signature scheme is existentially unforgeable under chosen message attacks (EU-CMA). Assume that commitment $Com_m'$ is securely constructed. Assume that \textsf{samples} are authenticated. Then protocol $\textbf{Prot}_{BM}$ securely realizes an ideal functionality $f$ which ensures model correctness in the TEE model for static adversaries.
\end{theorem}

Following the UC-style proof, we construct an ideal adversary $\textsf{S}$ against the ideal function $f$ (i.e., a simulator) in the idea model $\textsf{IDEAL}$ such that the environment $\mathcal{Z}$ as a distinguisher cannot distinguish the distribution in \textsf{IDEAL} from that in the $f_{\textsf{TEE}}$-hybrid model $f_{\textsf{TEE}}$-\textsf{HYBRID}, where a static adversary $\mathcal{A}$ interacts with protocol $\textbf{Prot}_{BM}$:
$$
\{\textsf{IDEAL}_{f,\textsf{S},\mathcal{Z}}\}_{\lambda}\\
\equiv_{c}\\
\{f_{\textsf{TEE}}\textsf{-HYBRID}_{\textbf{Prot}_{BM},\mathcal{A},\mathcal{Z}}\}_{\lambda},
$$
where $\equiv_{c}$ means that two distributions are computationally indistinguishable, and $\lambda$ is the security parameter.

Considering seller is malicious, we prove the computational indistinguishability above by sequentially constructing six hybrids.
Specifically, simulation \textsf{S} emulates the view of $\mathcal{A}$ in hybrids one by one, starting from the real execution of protocol $\textbf{Prot}_{BM}$, which makes the distinguisher $\mathcal{Z}$ fail to distinguish the distribution between hybrids.
Six hybrids are as follows: (a) In hybrid $\mathcal{H}_1$, \textsf{S} perfectly emulates the behaviors of TEE, and others follow the real protocol; (b) In hybrid $\mathcal{H}_2$, $\textsf{S}$ proceeds as $\mathcal{H}_1$ and it will abort, when receiving any message-signature tuple which is not generated by the TEE; (c) In hybrid $\mathcal{H}_3$, $\textsf{S}$ proceeds as $\mathcal{H}_2$ and it will abort, when receiving a commitment whose committed value is not pre-stored inside the TEE; d) In hybrid $\mathcal{H}_4$, $\textsf{S}$ proceeds as $\mathcal{H}_3$ and it will abort, when receiving a message-signature tuple which is sent by an enclave with different identifiers; e) In hybrid $\mathcal{H}_5$, $\textsf{S}$ proceeds as $\mathcal{H}_4$ and it will abort, when inspecting that $\mathcal{A}$ forwards incorrect messages from the TEE to \textsf{BM-Contract}; f) In hybrid $\mathcal{H}_6$, \textsf{S} emulates the messages that $\mathcal{A}$ interacts with $\textbf{Prot}_{BM}$ and then interacts with the ideal function $f$. With the last hybrid, \textsf{S} in \textsf{IDEAL} can faithfully emulate the view of the static adversary $\mathcal{A}$ in $f_{\textsf{TEE}}$-\textsf{HYBRID}. More proof details about the case of honest seller can be referred to the proof for work~\cite{tramer2017sealed}'s Theorem 1.

When considering the case of adaptive adversaries, where a malicious host can adaptively query the states of a enclave, our refinement design resists this kind of adversaries by introducing an instance of the ELI scheme~\cite{kaptchuk2019giving} in Section~\ref{sec:refinement}. The instance enables stateful multi-step execution and keeps the state freshness before each execution.
The ELI scheme is provably secure, since Lemma~\ref{them:theorem_third} has been proved in the appendix of work~\cite{kaptchuk2019giving}. Note that the inputs, i.e., the benchmark samples injected into the enclave can be in plaintext in our case. We thus need not use a commitment with the hiding property for the inputs, but they should be committed on the blockchain in advance.
Additionally, a seller entering the model marketplace should use a secure commitment mechanism to generate commitments to the model and the secret, that is, $Com_m$ and $Com_k$, respectively~\cite{goldreich2009foundations}. Meanwhile, the seller's off-line proof $p_k$ (see Section~\ref{subsec:preparation}) can be generated based on the SGPs protocol and their security have proved in work~\cite{tramer2017sealed}.\looseness=-1
\begin{theorem}\label{them:theorem_third}
Assume a secure authenticated encryption scheme $\textsf{Enc}(\cdot)$, a collision resistant hash function $\textsf{Hash}(\cdot)$, a secure pseudorandom function $\textsf{F}(\cdot)$, and the underlying blockchain is unforgeable. Then the ELI scheme is simulation secure.
\end{theorem}

\subsection{Exchanging fairness}
The model-money swapping process is fair by leveraging SGPs protocol and BE-Contract. A buyer cannot reject to pay fees, once he receives $\textsf{AEnc}_{pk_{\textcolor[rgb]{0.5,0.1,0.1}{\mathcal{B}}}}$ ($k_{m}$) which is proved correctly generated, since his/her deposit on the BE-Contract can be automatically sent to the seller's account. On the other hand, the seller cannot gain the payment from the buyer only if he does not reveal $\textsf{AEnc}_{pk_{\textcolor[rgb]{0.5,0.1,0.1}{\mathcal{B}}}}$ ($k_{m}$) in time, since the buyer's deposit can be refunded. We present the security analysis for protocol $\textbf{Prot}_{ME}$ in Lemma~\ref{them:theorem_fourth}. More details refer to the security proof of work\cite{tramer2017sealed}'s Theorem~3.

\begin{theorem}\label{them:theorem_fourth}
Assume that the used TEE's signature scheme is existentially unforgeable under chosen message attacks (EU-CMA). Assume that $\textsf{AEnc}(\cdot)$ is semantic secure. Then protocol $\textbf{Prot}_{ME}$ securely realizes an ideal functionality $g$ combining an SGP and the BE-Contract which ensures fairness for static adversaries.
\end{theorem}

A simulator $\textsf{S}$ is constructed against the ideal function $g$ in the idea model $\textsf{IDEAL}$ such that the environment $\mathcal{Z}$ cannot distinguish the distribution in \textsf{IDEAL} from that in the $g_{\textsf{SGP}}$-hybird model $g_{\textsf{SGP}}$-\textsf{HYBRID}, where a static adversary $\mathcal{A}$ interacts with protocol $\textbf{Prot}_{ME}$:
$$
\{\textsf{IDEAL}_{g,\textsf{S},\mathcal{Z}}\}_{\lambda}\\
\equiv_{c}\\
\{g_{\textsf{SGP}}\textsf{-HYBRID}_{\textbf{Prot}_{ME},\mathcal{A},\mathcal{Z}}\}_{\lambda},
$$
where $\equiv_{c}$ means that two distributions are computationally indistinguishable, and $\lambda$ is the security parameter.

With the proceeding round in the blockchain setting, \textsf{S} emulates messages between $\mathcal{A}$ and $\textbf{Prot}_{ME}$ to interact with $g$ via four hybrids. Starting from the execution of the real protocol, the distributions between two successive hybrids are distinguishable: a) In hybrid $\mathcal{H}_1$, \textsf{S} perfectly emulates the behaviors of the functionality of an SGP and the BE-Contract, and others follow the execution in the real world; b) In hybrid $\mathcal{H}_2$, \textsf{S} can perfectly emulate the `Init' phase to initialize the BE-Contract and the `commit' request sent to the SGP. Note that the indistinguishability between $\mathcal{H}_2$ and $\mathcal{H}_1$ is based on the security definition of the functionality of the SGP; c) In hybrid $\mathcal{H}_3$, \textsf{S} uses the randomness which is sampled by the ideal contract in the SGP, instead of the randomness sampled by himself in $\mathcal{H}_2$; d) In hybrid $\mathcal{H}_4$, \textsf{S} replaces $\textsf{AEnc}_{pk_{\textcolor[rgb]{0.5,0.1,0.1}{\mathcal{B}}}}$ ($k_{m}$) with $\textsf{AEnc}_{pk_{\textcolor[rgb]{0.5,0.1,0.1}{\mathcal{B}}}}$ ($0$) sent to the BE-Contract. Note that the indistinguishability between $\mathcal{H}_4$ and $\mathcal{H}_3$ is  based on the semantic security of $\textsf{AEnc}$.

\section{\wjss{PERFORMANCE ANALYSIS}}
\wjss{We proceed to present the theoretic performance analysis of each entity in our design including the benchmarking and monetization stages.
Concretely, we analyse the performance costs in terms of space, computational and communication complexity.}

\noindent\textbf{\wjss{Benchmarking Stage:}}
\wjss{Here we theoretically analyse the performance costs of the \textsf{BM-Contract} and a seller participating in this stage.
%
%
Note that the symbol $|a|$ in the following means the size of element $a$.}

\wjss{In terms of the \textsf{BM-Contract}, the space complexity is $4\times256$bits+$|\textcolor[rgb]{0,1,0}{prog}|+|\textsf{outp}|$, since the contract needs to store \textcolor[rgb]{0,0,1}{${Addr}_{m}$}, $ID_m$, $Com_m$, $\textsf{Hash(samples)}$ (the total sizes of the four elements are $4\times256$bits), \textcolor[rgb]{0,1,0}{$prog$} and \textsf{outp}.
The computational complexity mainly comes from one hash operation.
Lastly, the communication complexity is $|\textcolor[rgb]{0,1,0}{prog}|+2\times 256$bits$+2\times 70$bytes$+3\times 70$bytes$+|\textsf{outp}|$.
Herein, $2\times 256$ bits describe the sizes of $Com_m$ and $\textsf{Hash(samples)}$ in total;
$2\times 70$bytes are the sizes of $\sigma_{\textcolor[rgb]{0.5,0.1,0.1}{\mathcal{I}}}$ and $\sigma_{att}$;
another $2\times 70$bytes are the total sizes of $tx_{c}^{\textcolor[rgb]{0.5,0.1,0.1}{\mathcal{E}}}$ and $tx_{o}^{\textcolor[rgb]{0.5,0.1,0.1}{\mathcal{E}}}$.}

\wjss{In terms of a participating seller, the space complexity is $256$bits$+ |\textsf{samples}|+|\textsf{model}|+|\textcolor[rgb]{0,1,0}{prog}|+|\textsf{outp}|$, $256$bits mean the size of $Com_m$.
The computational complexity contains the operations of installing program, computing commitment, verifying attestation and evaluating program.
Next, the communication complexity is $|\textcolor[rgb]{0,1,0}{prog}|+2\times 256$bits $+2\times 70$bytes$+70$bytes$+|\textsf{model}|+|\textsf{samples}|+|\textsf{outp}|$.
To be specific, $2\times 256$ describe the sizes of $Com_m$ and $\textsf{Hash(samples)}$;
$2\times 70$bytes are the total sizes of $tx_{c}^{\textcolor[rgb]{0.5,0.1,0.1}{\mathcal{E}}}$ and $tx_{o}^{\textcolor[rgb]{0.5,0.1,0.1}{\mathcal{E}}}$.}

\noindent\textbf{\wjss{Monetization Stage:}}
\wjss{We now proceed to introduce the performance costs during the monetization stage, involving three parties including a seller, a buyer and the \textsf{BE-Contract}.}

\wjss{For the seller, the space complexity is $2\times256$bits$+64$bytes$+|\textsf{AEnc}_{pk_{\textcolor[rgb]{0.5,0.1,0.1}{\mathcal{B}}}}(k_m)|+70$bytes.
Here, $2\times256$bits are the sizes of $Com_k$ and $k_m$;
$64$bytes represent the size of $pk_{\textcolor[rgb]{0.5,0.1,0.1}{\mathcal{B}}}$; another $70$bytes refer to the size of $\sigma_{o}^{'}$.
The computational complexity comes from the operations of computing commitment $Com_k$ and generating ciphertext of $k_m$.
Lastly, the communication complexity is $64$bytes$+2\times256$bits$+70$bytes$+(64$bytes$+|\textsf{AEnc}_{pk_{\textcolor[rgb]{0.5,0.1,0.1}{\mathcal{B}}}}(k_m)|)$.
Concretely, the seller fetches $pk_{\textcolor[rgb]{0.5,0.1,0.1}{\mathcal{B}}}$ with $64$bytes and $Com_k$ with $256$bits from the \textsf{BE-Contract};
from its enclave, the seller receives $\sigma_{o}^{'}$, $pk_{\textcolor[rgb]{0.5,0.1,0.1}{\mathcal{B}}}$ and $\textsf{AEnc}_{pk_{\textcolor[rgb]{0.5,0.1,0.1}{\mathcal{B}}}}(k_m)$ which are $70$bytes$+(64$bytes$+|\textsf{AEnc}_{pk_{\textcolor[rgb]{0.5,0.1,0.1}{\mathcal{B}}}}(k_m)|)$.}

\wjss{In terms of the buyer, we mainly discuss the computational and communication costs.
From the aspect of computational complexity, the buyer executes two decryption operations for decrypting $\textsf{AEnc}_{pk_{\textcolor[rgb]{0.5,0.1,0.1}{\mathcal{B}}}}(k_m)$ with his private key and successively decrypting the encrypted model with $k_m$.
From the aspect of communication complexity, the buyer needs to submit $Com_k$, $ID_m$, $pk_{\textcolor[rgb]{0.5,0.1,0.1}{\mathcal{B}}}$ and $p_k$ which correspondingly are $256$bits, $256$bits, $64$bytes and $70$bytes.}

\wjss{In terms of the \textsf{BE-Contract}, the space complexity is $2\times256$bits$+64$bytes$+|\textsf{AEnc}_{pk_{\textcolor[rgb]{0.5,0.1,0.1}{\mathcal{B}}}}(k_m)|+2\times70$bytes, since the contract stores $Com_k$ and $ID_m$ which are $2\times256$bits, $pk_{\textcolor[rgb]{0.5,0.1,0.1}{\mathcal{B}}}$ which is $64$bytes, as well as $p_k$ and $\sigma_o^{'}$ which are $2\times70$bytes.
As for the computational complexity, it refers to the operation of verifying attestation $\sigma_o^{'}$.
The last one, communication complexity comes from $Com_m$, $ID_m$, $pk_{\textcolor[rgb]{0.5,0.1,0.1}{\mathcal{B}}}$ and $p_k$ (sent by the buyer) which are $64$bytes$+2\times256$bits$+70$bytes, and $\sigma_{o}^{'}$, $pk_{\textcolor[rgb]{0.5,0.1,0.1}{\mathcal{B}}}$ and $\textsf{AEnc}_{pk_{\textcolor[rgb]{0.5,0.1,0.1}{\mathcal{B}}}}(k_m)$ which are $70$bytes$+(64$bytes$+|\textsf{AEnc}_{pk_{\textcolor[rgb]{0.5,0.1,0.1}{\mathcal{B}}}}(k_m)|)$ (sent by the seller).}
\section{EXPERIMENTS}
In this section, we introduce a prototype implementation and conduct a series of experiments using the standard benchmark datasets.
\subsection{Implementation}
\noindent\textbf{Model Benchmarking with TEE.} We use the Intel's SGX as the TEE, since it is widely used. We also employ a memory-safe and lightweight library operating system (LibOS) for SGX, called as Occlum\footnote{https://github.com/occlum/occlum} to deploy models in the enclave, through which hundreds of lines of SGX-aware codes need not be written. This way still supports attesting the codes running inside the enclave only if we put a Service Provider ID (SPID) and the associated certificate in the appointed file path, which enables Occlum to access Intel Attestation Service (IAS).\looseness=-1

We utilize popular pre-trained models on TensorFlow Lite\footnote{https://www.tensorflow.org/lite/guide/hosted\_models}, such as Mobilenet\_V1\_1.0\_224  (MobilenetV1 for short), Mobilenet\_V2\_1.0\_224 (MobilenetV2), NASNet mobile, ResNet\_V2\_101 (ResNet101) and Inception\_V3. Their information are shown in TABLE~\ref{tab:models}, in which top-1 errors for each model are estimated on ImageNet classification. The reason we utilize TensorFlow Lite models is due to the memory limitation of the current version SGX enclave, only about $93$~MB available. In our future work, we would explore new techniques, \emph{e.g.}, TVM~\cite{chen2018tvm}, to run more generic models inside the enclave.
\begin{table}[htbp]
 \centering
 \caption{\label{tab:models} Pre-trained models on TensorFlow Lite.}
 \begin{tabular}{lcc}
  \toprule
  Model & Size & Top-1 Error Rate \\
  \midrule
  MobilenetV1 & $16.9$~MB & $29.0\%$    \\
  MobilenetV2 & $14.0$~MB & $28.2\%$    \\
  NASNet mobile & $21.4$~MB & $26.1\%$    \\
  Inception\_V3 & $95.3$~MB & $22.1\%$    \\
  ResNet101 & $178.3$~MB & $23.2\%$    \\
  \bottomrule
 \end{tabular}
  \vspace{-10pt}
\end{table}

\noindent\textbf{Smart Contract.} We implement the BM-Contract and the BE-Contract with the Solidity programming language of Ethereum~\cite{Ethereum}. We deploy the two contracts to the Ethereum Test Network provided by MetaMask~\footnote{https://metamask.io/}.\looseness=-1
\subsection{Setup and Evaluation}
\noindent\textbf{Benchmark Datasets Setup.}
We download ImageNet-C, ImageNet-P\footnote{https://github.com/hendrycks/robustness} and ImageNet validation dataset\footnote{http://www.image-net.org/challenges/LSVRC/2012/} following the instruction of work~\cite{hendrycks2019benchmarking} and construct three benchmark samples. Concretely, ImageNet-C and ImageNet-P are built from the ImageNet validation dataset which consists of $50,000$ 299x299 images with $1000$ classes, added by corruptions and perturbations, respectively. The corruptions and perturbations mainly include four categories, such as digital, noise, weather and blur. ImageNet-C has $15$ kinds of corruptions and each corruption contains $5$ severity levels. They are all applied on the ImageNet validation dataset. Similarly, ImageNet-P has $10$ types of perturbation sequences and each sequence contains $31$ frames. Each sequence begins with a clean image (\emph{i.e.} the first frame) being ingested perturbation slightly and an intermediate frame is a perturbed frame of the previous one. The format of sequences is MP4 not JEPG. We consistently convert the format of ImageNet-C images and ImageNet-P sequence frames into the format of bmp, before inputting them into a model.~\looseness=-1

Three benchmark samples, $Set_1$, $Set_2$ and $Set_3$ then are randomly chosen from ImageNet-C, ImageNet-P and ImageNet validation dataset, respectively. $Set_1$ totally includes $15\times5\times100$ images, which is constructed by randomly choosing an image with respective to each severity level per corruption among multiple classes. $Set_2$ includes $10\times31\times100$ frames which are randomly chosen with respect to each perturbation sequence covering multiple classes. Last, $Set_3$ consists of $100$ images which are randomly chosen per class.

In addition, we conduct the model evaluation experiments inside a SGX on an Ubuntu $18.04$ server with a 4-core Intel i5-7500 CPU $3.40$~GHz processor and $32$~GB RAM.

\noindent\textbf{Model Performance Metrics.}
We show two robustness metrics and the clean error of pre-trained models in TABLE~\ref{tab:metrics}. They are evaluated on benchmark samples $Set_1$, $Set_2$ and $Set_3$, respectively. Like work~\cite{hendrycks2019benchmarking}, we select AlexNet to be a baseline, and the robustness metrics of AlexNet are used to standardize the respective metrics of the benchmarked models.
Robustness metrics include \textsf{mCE} and \textsf{mFP}, as well as accuracy metric is clean error \textsf{ce}.
\textsf{mCE} and \textsf{mFP} are calculated guided by the formulas in work~\cite{hendrycks2019benchmarking}.

\begin{figure}[!htb]
\vspace{-5pt}
\begin{tabular}{cc}
\begin{minipage}[t]{0.48\linewidth}
    \includegraphics[width = 1\linewidth]{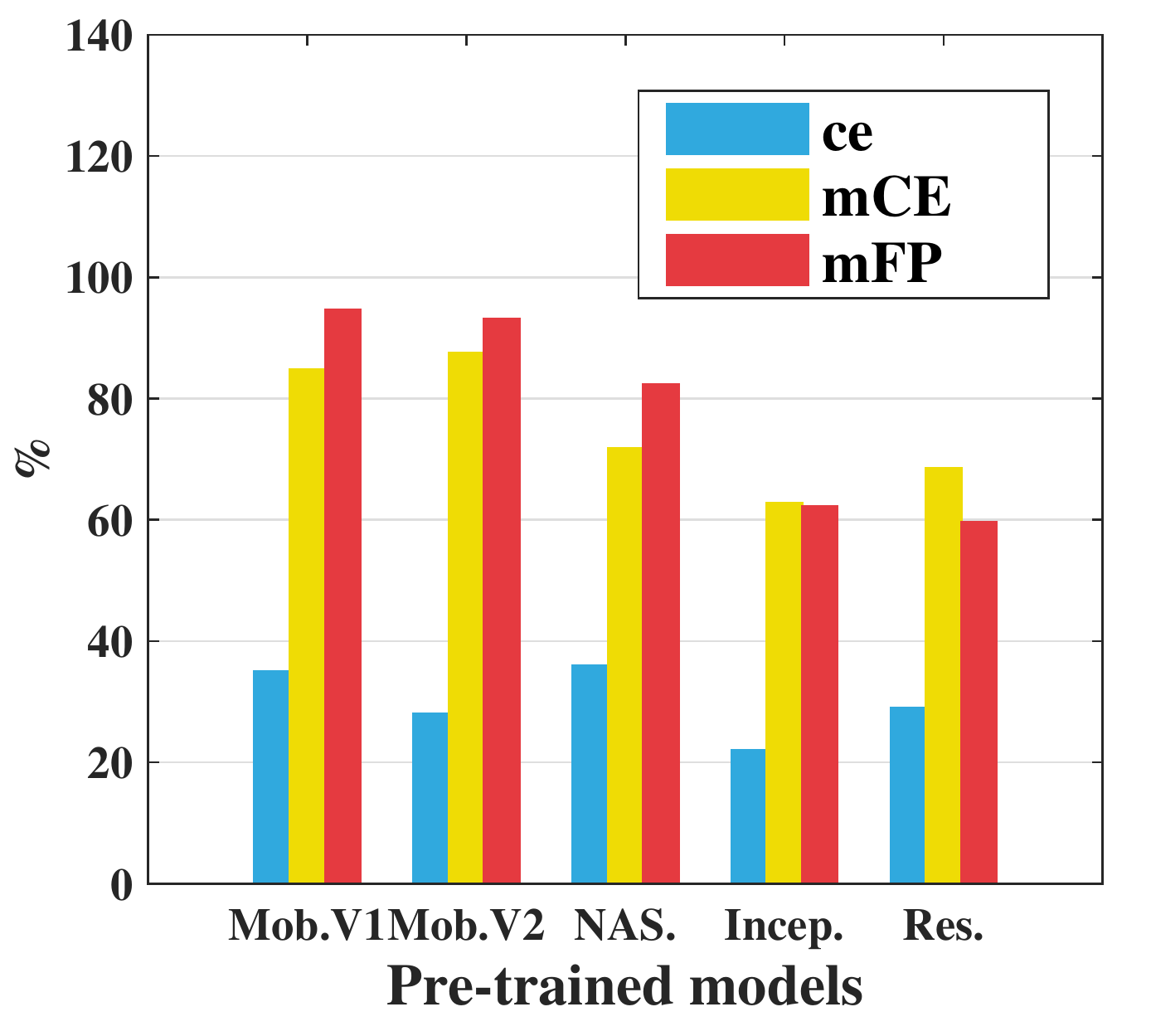}
            \caption{mCE and mFP.}\label{fig:mCE_and_mFP}
\end{minipage}
\begin{minipage}[t]{0.48\linewidth}
    \includegraphics[width = 1\linewidth]{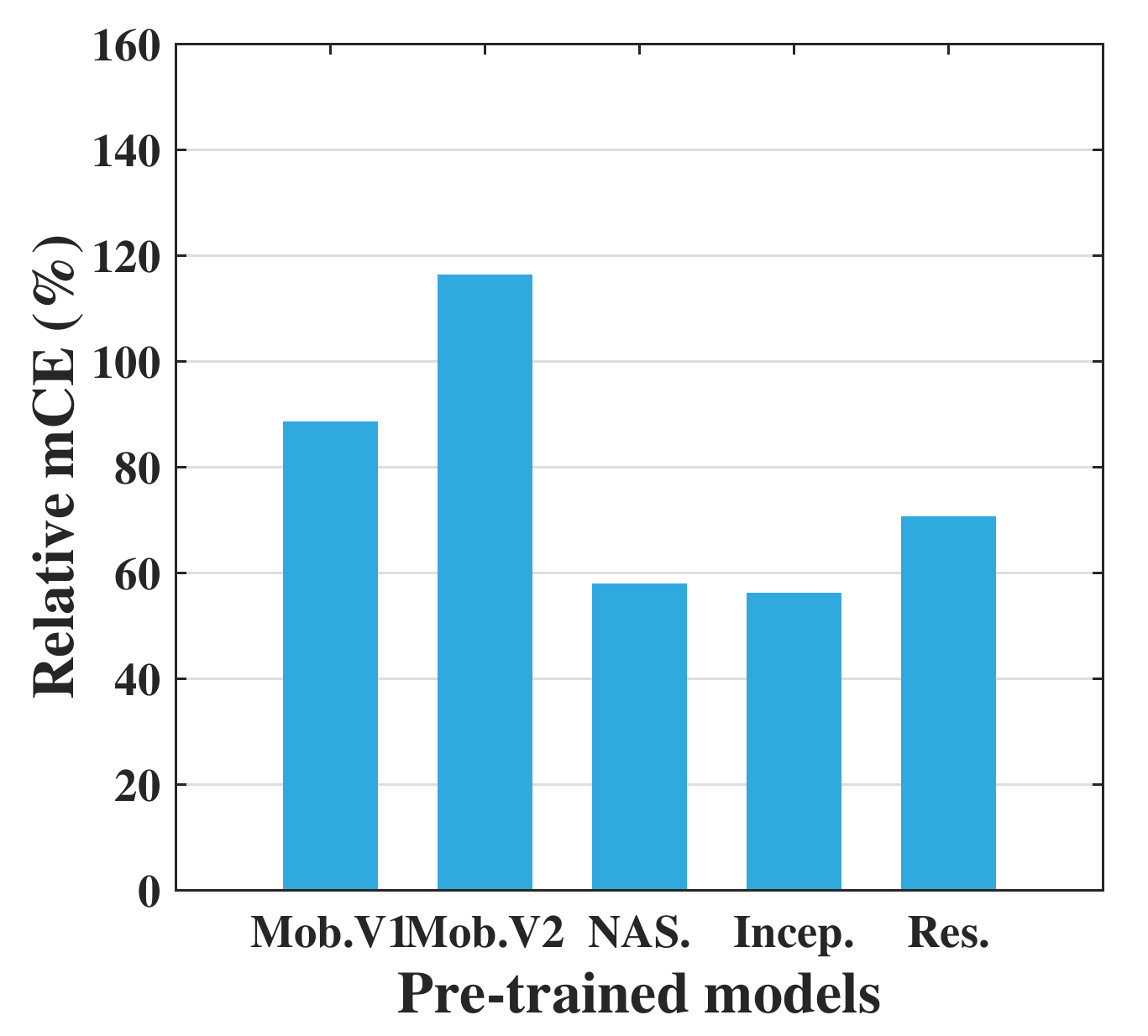}
            \caption{Relative mCE.}\label{fig:rmCE_comp}
\end{minipage}
\end{tabular}
\vspace{-10pt}
\end{figure}

Fig.~\ref{fig:mCE_and_mFP} shows that Inception\_V3 owns the stronger corruption and perturbation robustness than others as it has the lowest \textsf{mCE} and \textsf{mFP}. From TABLE~\ref{tab:metrics}, we can see that each of models have the higher \textsf{mCE} and \textsf{mFP} than the corresponding \textsf{ce}. That means that these models can perform poorly when meeting corrupted data and perturbed data. Particularly, MobilenetV2 degrades most obviously among all presented models on corrupted data, since it should have a comparably lower clean error $28.0\%$ but shows the highest \textsf{mCE} $87.6\%$, which indicates it has the worst corruption robustness.

\begin{table}[htbp]
\vspace{-5pt}
 \centering
 \caption{\label{tab:metrics} Performance metrics of pre-trained models.}
 \begin{tabular}{lccc}
  \toprule
  Model & \textsf{ce} & \textsf{mCE}  & \textsf{mFP} \\
  \midrule
  AlexNet & $43.5\%$ & $100.0\%$ & $100.0\%$\\
  MobilenetV1 & $35.0\%$ & $84.8\%$ & $94.6\%$\\
  MobilenetV2 & $28.0\%$ & $87.6\%$  & $93.2\%$\\
  NASNet mobile & $36.0\%$ & $71.8\%$  & $82.4\%$\\
  Inception\_V3 & $22.0\%$ & $62.8\%$  & $62.2\%$\\
  ResNet101 & $29.0\%$ & $68.5\%$  & $59.6\%$\\
  \bottomrule
 \end{tabular}
 \vspace{-10pt}
\end{table}
To further describe this observation, we compute another metric, \emph{i.e.}, the relative mCE. It measures the gap between the \textsf{mCE} and the \textsf{ce}, describing a model's accuracy degradation on the corrupted data. The above observation can be further verified by Fig.~\ref{fig:rmCE_comp}. Concretely, MobilenetV2 has the top relative mCE $118.9\%$ among all models while Inception\_V3 has the lowest one $56.0\%$. In this case, a buyer who wants to obtain a model with corruption robustness can purchase Inception\_V3 among those benchmarked models. On the other hand, although there exists no a relative metric for \textsf{mFP}, a buyer can choose models with the higher perturbation robustness according to a lower \textsf{mFP}, \emph{e.g.}, ResNet101.

\begin{figure}[!htb]
\vspace{-5pt}
\begin{tabular}{cc}
\begin{minipage}[t]{0.48\linewidth}
    \includegraphics[width = 1\linewidth]{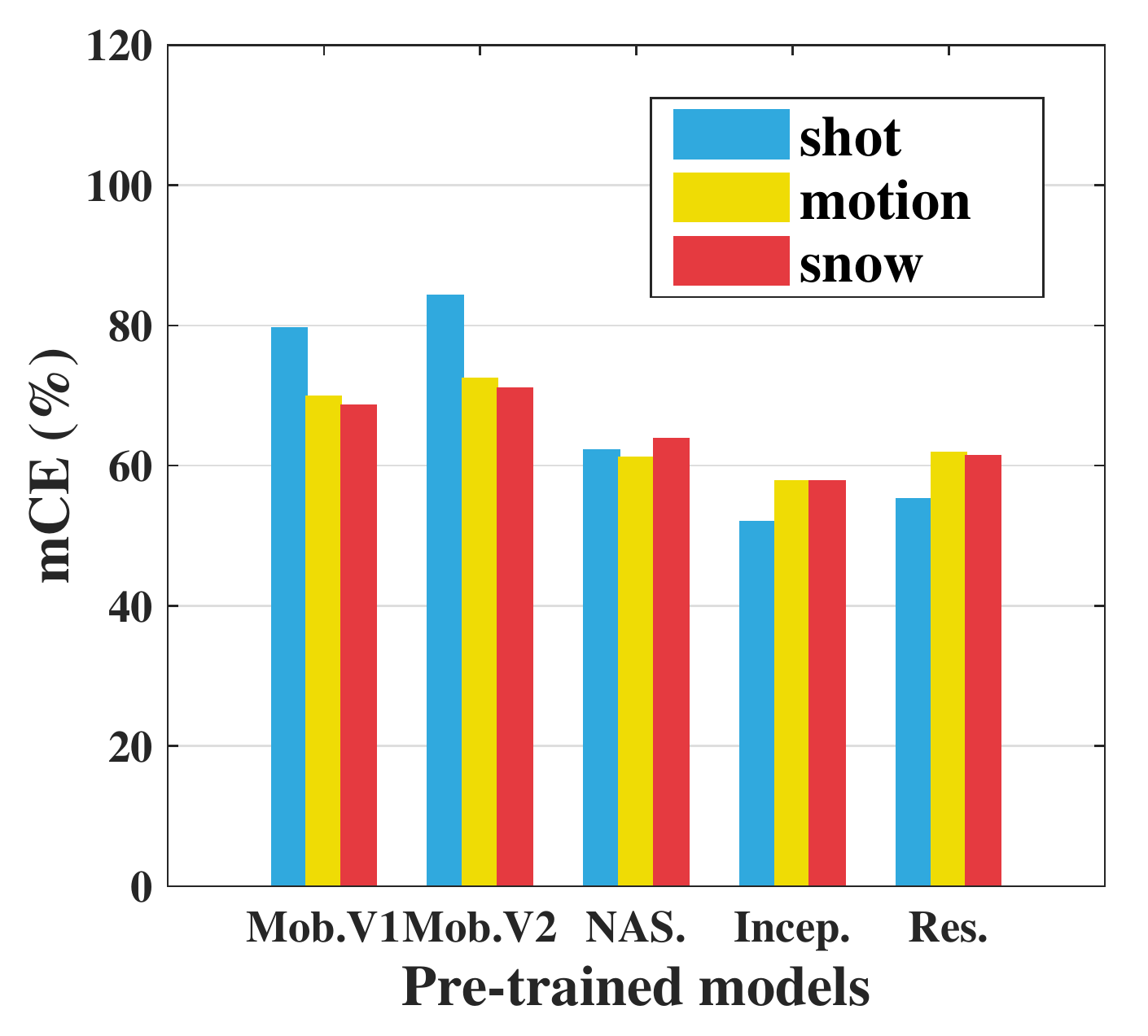}
    \caption{mCE on corruption shot, motion and snow.}\label{fig:mCE}
\end{minipage}
\begin{minipage}[t]{0.48\linewidth}
    \includegraphics[width = 1\linewidth]{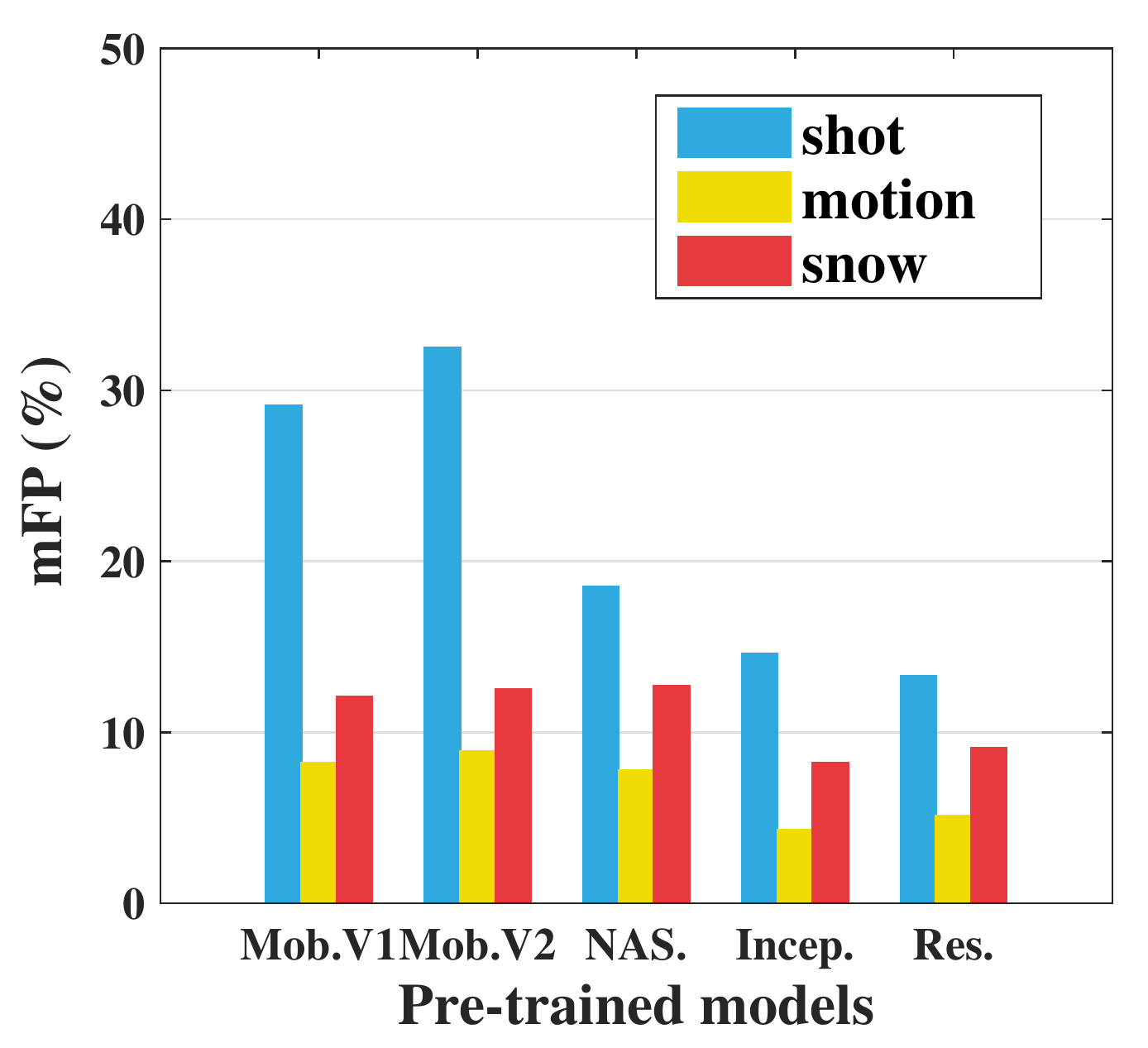}
    \caption{mFP on perturbation shot, motion and snow.}\label{fig:mFP}
\end{minipage}
\end{tabular}
\vspace{-10pt}
\end{figure}

From TABLE~\ref{tab:metrics}, Inception\_V3 and RestNet101 are more robust than the remaining models, since their \textsf{mCE} and \textsf{mFP} are comparably lower. For giving a fine-grained view, we show Fig.~\ref{fig:mCE} and Fig.~\ref{fig:mFP}, where the \textsf{mCE} and \textsf{mFP} are estimated on three categories of corruptions and perturbations (\emph{i.e.}, shot, motion and snow), respectively. It is clear that Inception\_V3 and ResNet101 have a better robustness on each category, which is consistent to the statistical data given in TABLE~\ref{tab:metrics}.

\noindent\textbf{Computational Costs and Complexity.}
We estimate the computational costs of a seller and a buyer accordingly, considering the expends for a seller entering the market and a buyer using our market.

First of all, a seller on one hand needs to upload his model information to the blockchain via transactions, which causes gas costs $\textsf{Gas}_s$. Concretely, the most costs of sending transactions come from uploading proofs $p_k$, $p_{ck}$, $p_{cm}$ and commitments $Com_k$, $Com_m$ as described in Section \ref{subsec:preparation}. Proofs can be the attestations generated by the trusted hardware with the Elliptic Curve Digital Signature Algorithm, which are totally $210$~bytes as each one is $70$~bytes. Commitments can be implemented by using Pedersen commitment algorithm. For reducing the sizes of commitments, we further use a hash function, \emph{i.e.}, SHA-256, to generate the hash values of each commitment, in which each hash value is $256$~bits. Thus, the seller expends about $\textsf{Gas}_s$ $57,174$ units (about $0.000057$~ETH) to upload the entire model information.
On the other hand, the seller has computation cost $\textsf{Comp}_s$ of benchmarking his model.

\begin{figure}[htbp]
    \centering
    \includegraphics[width=5cm]{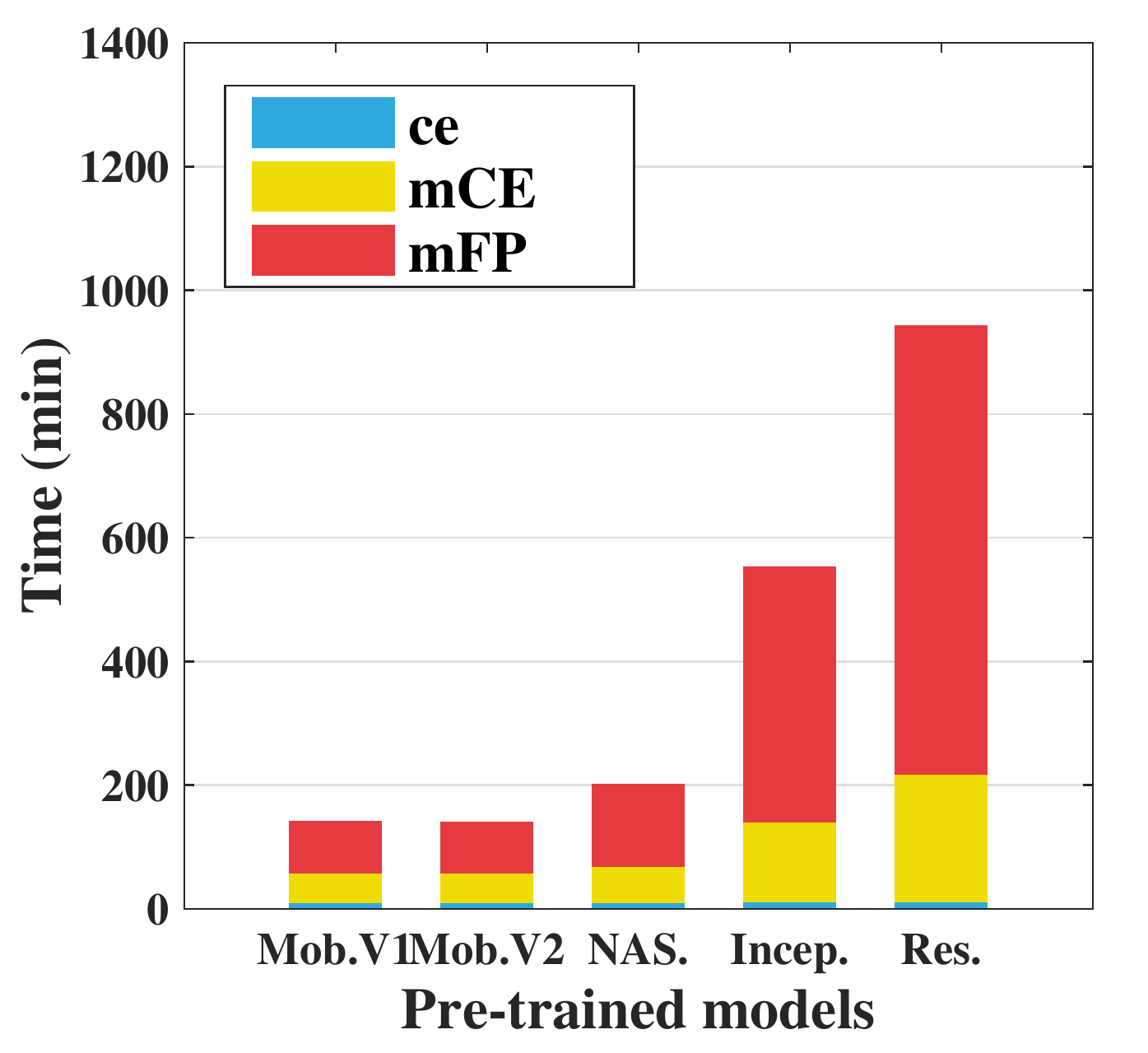}
    \caption{Time complexity for evaluating each metrics.}\label{fig:time}
     \vspace{-10pt}
\end{figure}

Now, we explain $\textsf{Comp}_s$ in terms of time complexity with respect to different-size models as shown in TABLE~\ref{tab:times}. For the column of total time, each row is the total consuming times for benchmarking on samples $Set_1$, $Set_2$ and $Set_3$ (the total size is $38,600$). Also, the separate time costs for evaluating each metric are presented in Fig.~\ref{fig:time}. In detail, a large-size model, \emph{e.g.}, Inception\_V3 and RestNet101 could take the sellers more time, but this may deserve, since the large-size model performs a comparably lower classification error, and buyers may prefer to purchase them.
But the average time for each query image is also crucial for an online MLaaS system, we would explore this concern in our future work.
\begin{table}[htbp]
 \centering
 \caption{\label{tab:times} Time complexity of benchmarking models.}
 \begin{tabular}{lccc}
  \toprule
  Model & Size & Total Time (min) & Avg. Time (s) \\
  \midrule
  MobilenetV1 & $16.9$~Mb & $130.27$  & $0.20$  \\
  MobilenetV2 & $14.0$~Mb & $129.95 $ & $0.20$  \\
  NASNet mobile & $21.4$~Mb & $190.36$ & $0.30$ \\
  Inception\_V3 & $95.3$~Mb & $541.96$  & $0.84$  \\
  ResNet101 & $178.3$~Mb & $931.30$ & $1.45$   \\
  \bottomrule
 \end{tabular}
 \vspace{-10pt}
\end{table}

\vspace{-5pt}
Secondly, a buyer needs to initiate and deploy BE-Contract interacting with the seller and thus spends gas costs $\textsf{Gas}_b$ for it. Recall that BE-Contract defines three entry points: \textsf{Init}, \textsf{Request} and \textsf{Publish}. After deploying this contract with a deposit, the buyer would execute the defined logics accordingly every time each entry point is invoked. During a simulated interaction, $\textsf{Gas}_b$ needs $1,206,886$~units gas, about $0.001207$~ETH, in which $981,901$~units for \textsf{Init}, $27,873$~units for \textsf{Request} and $197,112$~units for \textsf{Publish}.

\section{DISCUSSION}\label{sec:discuss}
\noindent\textbf{Beyond object recognition tasks.}
\wjss{This work mainly considers object recognition tasks~\cite{tractica9AI}.
Despite this, our model marketplace can be extended to other two popular recognition services, \emph{i.e.}, speech recognition and text recognition.
Extending to such services needs to replace the benchmark datasets used in this work with other standard datasets specialized for speech recognition or text recognition.
Available benchmark datasets, such as speech commands dataset~\cite{warden2017speech} for speech recognition and SQuAD~\cite{jia2017adversarial} for text recognition can be used.
Besides, crafting adversarial examples for these two groups are undergoing investigation~\cite{kuleshov2018adversarial, carlini2018audio}.
We envision that the future standardized benchmark datasets help reintegrating the component of benchmarking models.}
%

\noindent\wjss{\textbf{Mitigating the limitations of Intel's SGX.}}
Here we figure out SGX's limitations from the following four aspects and discuss the countermeasures corresponding to the limitations:
\wjss{\emph{(a) handling the memory limit.}
In the current work, we run the models whose sizes exceed the inherit memory limit of SGX by leveraging Linux's paging technique which supports using more protected memory.
But it is well-known that paging is inefficient~\cite{orenbach2017eleos, orenbach2019cosmix}.
In our future work, we will explore the off-the-shelf optimization technologies like application-managed paging~\cite{orenbach2017eleos, orenbach2019cosmix} to reduce the high overhead of paging.}
\wjss{\emph{(b) setting up fault-tolerant TEE-empowered components.}
Recall that we implement an off-chain oracle for authentically relaying outside-blockchain data with a centralized TEE-empowered component, by following a common practice like the existing work~\cite{zhang2016town}.
Nevertheless, it causes the single-point-of-failure concern.
Once the component is compromised or the TEE's secret key is stolen, data authenticity is hard to be guaranteed.
To resolve this issue, our natural idea is to make the centralized component decentralized.
Concretely speaking, we can leverage multiple TEE-empowered components instead of the single one, and enable them to make Byzantine consensus on the sampling test set when requesting the outside benchmarking data.
Besides, we can make the secret key inside the TEE securely managed by a group of decentralized nodes, so as to against the threat of compromising the TEE's secret key~\cite{das2019fastkitten}.}
\wjss{\emph{(c) achieving confidential attestation.}
In the previous section, we assume that program $\textcolor[rgb]{0,1,0}{{Prog_m}}$ representing the architecture of a model does not contain the model's private information, since we do not make the model's parameters public.
This assumption may be somewhat strong.
We aware that Liu \emph{et al.}\cite{liu2020confidential} present a confidential attestation model, and believe that it might pave the way for exploring the approach to protecting the confidentiality of program $\textcolor[rgb]{0,1,0}{{Prog_m}}$.}
\wjss{\emph{(d) enabling distributed and frequent attestation.}
Our design regards attestations built by an enclave as the transactions on the blockchain, and they will be verified by the nodes on the blockchain according to the transaction verification protocol~\cite{zhang2016town}.
It means that the distributed nodes on the blockchain will frequently request IAS for verifying the attestations.
Such distributed and frequent requests could lead to a high overhead on IAS, causing a threat of single point of failure.
Aiming to resolve this issue, Chen \emph{et al.}\cite{chen2019opera} propose an open remote attestation platform.}

\noindent\wjss{\textbf{Defenses against dataset extraction attack vector.}}
\wjss{This attack vector describes that a malicious seller might commit to multiple models for repeating the benchmark process, in an effort to extract the entire benchmark dataset.
We discuss several defense approaches against the attack vector:
%
%
(a) \emph{updating the benchmark datasets periodically}. New samples can be added to certain benchmark dataset at each sampling period.
\wjst{(b) \emph{protecting the test samples during model evaluation}. Two primary directions for test samples protection can be considered; one is encrypting the samples and another one is adding noises into the samples. Specifically, we can encrypt the test samples by using a data provider's public key generated by a homomorphic encryption algorithm. With the homomorphic property of the encryption algorithm, model evaluation on the encrypted test samples can be securely executed without revealing the test samples. Besides the encryption, we can also add noises into the test samples, so as to conceal the test samples.
(c) \emph{enlarging the test sample space via randomly selected data transformation}. It starts with preparing various data transformation types, e.g., rotation and flip, etc. in advance. Next, we can randomly chose one/multiple data transformation type(s) along with each sampling request, and then modify the samples according to the chosen data transformation operation. As a result, we enlarge the space of test samples.}
}

\noindent\wjss{\textbf{Technical correctness of models benchmarking via random sampling.}}
%
%
\wjss{Using test data via random sampling can raise a concern, that is, whether or not it is fair to benchmark different models using different sampling test data.
We argue that this concern can be addressed from the following three aspects:
(a) we can use the same category of test data to benchmark the models for the same category of recognition tasks.
%
(b) the models for the same recognition tasks are not necessarily evaluated using test sets via different sampling, which means that we can benchmark a collection of models in a batch manner.
(c) we can sample representative test sets from the benchmark datasets, by adopting the existing sampling techniques~\cite{yang2018lightweight, gretton2012kernel, anderson1994two}, which ensures that any two sampling test sets have the same distribution.
Thus, despite that different models are evaluated over different test sets via separate sampling,
their performance results are comparable.}
\section{CONCLUSION}
We present a secure and decentralized model marketplace for MLaaS in this paper. 
We realize the fair model monetization procedure based on the blockchain, which enables model buyers, \emph{e.g.}, MLaaS providers, to fairly purchase models from model sellers, \emph{e.g.}, individual developers, eliminating the potential disputes.
By treating the blockchain-based procedure as a starting point, we benchmark selling models to obtain their authentic performance that would be used to monetize the models.
We guarantee the correctness of the model benchmarking, by leveraging trusted hardware-based proofs and smart contracts.
Besides, we envision that our model marketplace can facilitate secure sharing of well-trained models and promote the development of MLaaS.
We lastly implement a prototype of our marketplace on Ethereum blockchain, and conduct the extensive experiments with the standard benchmark datasets, which demonstrates the affordable performance of our marketplace.
\section*{ACKNOWLEDGE}
Jian Weng was partially supported by National Key Research and Development Plan of China under Grant No. 2020YFB1005600, National Natural Science Foundation of China under Grant Nos. 61825203, U1736203 and 61732021, Major Program of Guangdong Basic and Applied Research Project under Grant No. 2019B030302008, and Guangdong Provincial Science and Technology Project under Grant No. 2017B010111005.
Dr. Wang was supported in part by the Research Grants Council of Hong Kong under Grant CityU 11217819 and Grant CityU 11217620, by the Innovation and Technology Commission of Hong Kong under ITF Project ITS/145/19, and by the National Natural Science Foundation of China under Grant 61572412.

\bibliographystyle{IEEEtran}
\bibliography{references}

\begin{IEEEbiography}[{\includegraphics[width=1in,height=1.25in,clip,keepaspectratio]{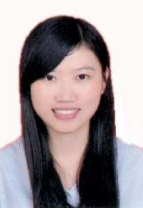}}]{Jiasi Weng} obtained the B.S degree in Software
engineering from South China Agriculture University
in June 2016. Currently, she is a
Ph.D. student with School of Information Science and Technology in Jinan University. Her research
interests include privacy-preserving machine learning and blockchain.
\vspace{-10pt}
\end{IEEEbiography}

\begin{IEEEbiography}[{\includegraphics[width=1.1in,height=1.25in,clip,keepaspectratio]{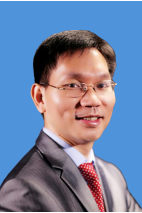}}]{Jian Weng}
is a professor and the Executive
Dean with College of Information Science and
Technology in Jinan University. He received B.S.
degree and M.S. degree from South China University of Technology in 2001 and 2004 respectively, and Ph.D. degree at Shanghai Jiao
Tong University in 2008. His research areas include public key cryptography, cloud security,
blockchain, etc. He has published 80 papers
in international conferences and journals such
as CRYPTO, EUROCRYPT, ASIACRYPT, TCC, PKC, CT-RSA, IEEE TPAMI, IEEE TDSC, etc.
He also serves as
associate editor of IEEE Transactions on Vehicular Technology. He
received the Young Scientists Fund of the National Natural Science
Foundation of China in 2018, and the Cryptography Innovation Award
from Chinese Association for Cryptologic Research (CACR) in 2015.
\end{IEEEbiography}

\begin{IEEEbiography}[{\includegraphics[width=1.1in,height=1.25in,clip,keepaspectratio]{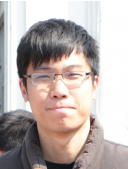}}]{Chengjun Cai}
received the BS degree in computer
science and technology from Jinan University, in 2016. He is working toward the PhD degree at the City University of Hong Kong. He was a research assistant with the City University of Hong Kong. His research interests include secure decentralized applications and privacy-enhancing technologies. He is a student member of the IEEE.
\end{IEEEbiography}

\begin{IEEEbiography}[{\includegraphics[width=1.1in,height=1.25in,clip,keepaspectratio]{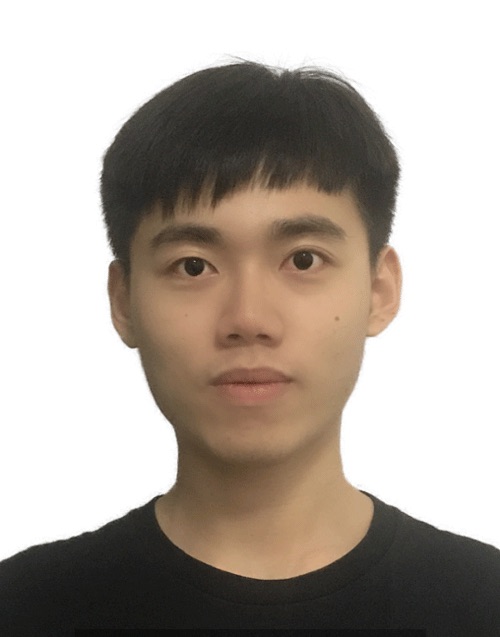}}]{Hongwei Huang} received the BS degree in software engineering from the South China Agriculture University (2014 - 2018). He has been pursuing his MS degree at Jinan University since 2018. His research interests include machine learning and its privacy and security.
\end{IEEEbiography}

\begin{IEEEbiography}[{\includegraphics[width=1.1in,height=1.25in,clip,keepaspectratio]{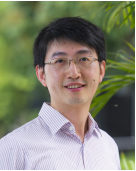}}]{Cong Wang} is an Associate Professor in the Department of Computer Science, City University of Hong Kong. His current research interests include data and network security, blockchain and decentralized applications, and privacy-enhancing technologies. He is one of the Founding Members of the Young Academy of Sciences of Hong Kong. He received the Outstanding Researcher Award (junior faculty) in 2019, the Outstanding Supervisor Award in 2017 and the President's Awards in 2019 and 2016, all from City University of Hong Kong. He is a co-recipient of the IEEE INFOCOM Test of Time Paper Award 2020, Best Paper Award of IEEE ICDCS 2020, Best Student Paper Award of IEEE ICDCS 2017, and the Best Paper Award of IEEE ICPADS 2018 and MSN 2015. He is a fellow of the IEEE, and member of the ACM.
\end{IEEEbiography}

\balance

\end{document}